\documentclass[usenatbib]{aa}
\usepackage{float,txfonts}
\usepackage[colorlinks]{hyperref}
\begin{document}

\newcommand{\pmill}{{\sc planck millennium~}}
\newcommand{\pms}{PMS~}
\newcommand{\lcdm}{$\Lambda$CDM }
\newcommand{\hmsun}{\, h^{-1}{\rm M}_\odot}
\newcommand{\msun}{\, {\rm M}_\odot}
\newcommand{\msunyr}{\, {\rm M}_\odot\ {\rm yr}^{-1}}
\newcommand{\gaea}{{\sc gaea~}}
\def\lesssim{\lower.5ex\hbox{$\; \buildrel < \over \sim \;$}}
\def\gtrsim{\lower.5ex\hbox{$\; \buildrel > \over \sim \;$}}

\title{Galaxy Assembly and Evolution in the P-Millennium simulation: galaxy clustering}
\author{Fabio Fontanot\inst{1,2}\fnmsep\thanks{e-mail:fabio.fontanot@inaf.it}
  \and Gabriella De Lucia\inst{1,2}
  \and Lizhi Xie\inst{3}
  \and Michaela Hirschmann\inst{4,1}
  \and Carlton Baugh\inst{5,6}
  \and John C. Helly\inst{5}
}
\institute{INAF - Astronomical Observatory of Trieste, via G.B. Tiepolo 11, I-34143 Trieste, Italy\label{inst1}
  \and IFPU - Institute for Fundamental Physics of the Universe, via Beirut 2, 34151, Trieste, Italy\label{inst2}
  \and Tianjin Astrophysics Center, Tianjin Normal University, Binshuixidao 393, 300384, Tianjin, China\label{inst4}
  \and EPFL - Institute for Physics, Laboratory for Galaxy Evolution, Observatoire de Sauverny, Chemin Pegasi 51, 1290 Versoix, CH\label{inst3}
  \and ICC - Institute for Computational Cosmology, Department of Physics, Durham University, South Road, Durham DH1 3LE, UK\label{inst5}
  \and Institute for Data Science, Durham University, South Road, Durham DH1 3LE, UK\label{inst6}
}

   \date{Received ???, 2024; accepted ???, 2024}

   \abstract{We present results from the latest version of the GAlaxy
     Evolution and Assembly (\gaea) theoretical model of galaxy
     formation coupled with merger trees extracted from the Planck
     Millennium Simulation (\pms). With respect to the Millennium
     Simulation, typically adopted in our previous work, the \pms
     provides a better mass resolution ($\sim$ 10$^8 \hmsun$), a
     larger volume (800$^3$ Mpc$^3$) and assumes cosmological
     parameters consistent with latest results from the Planck
     mission. The model includes, at the same time, a treatment for
     the partition of cold gas into atomic and molecular (H$_2$)
     components; a better treatment for environmental processes acting
     on satellite galaxies; an updated modelling of cold gas accretion
     on Super-Massive Black Holes, leading to the Active Galactic
     Nuclei (AGN) phenomenon and relative feedback on the host
     galaxy. We compare \gaea predictions based on the \pms, with
     model realizations based on other simulations in the Millennium
     Suite at different resolution, showing that the new model
     provides a remarkable consistency for most statistical properties
     of galaxy populations. We interpret this as due to the interplay
     between AGN feedback and H$_2$-based SFR (both acting as
     regulators of the cold gas content of model galaxies), as model
     versions considering only one of the two mechanisms do not show
     the same level of consistency. We then compare model predictions
     with available data for the galaxy 2-point correlation function
     (2pCF) in the redshift range 0<z$\lesssim$3. We show that \gaea
     runs correctly recover the main dependencies of the 2pCF as a
     function of stellar mass (M$_\star$), star formation activity,
     HI-content and redshift for M$_\star$ > 10$^{9}$ M$_\odot$
     galaxies. These results suggest that our model correctly captures
     both the distribution of galaxy populations in the Large Scale
     Structure and the interplay between the main physical processes
     regulating their baryonic content, both for central and satellite
     galaxies. The model predicts a small redshift evolution of the
     clustering amplitude, that results in an overprediction of
     z$\sim$3 clustering strength with respect to the available
     estimates, but is still consistent with data within 1-$\sigma$
     uncertainties.}


\titlerunning{Clustering in \gaea on \pms}\authorrunning{F.~Fontanot et al.}
   
   \maketitle

\section{Introduction}
\label{sec:intro}

The evolution of the large-scale spatial distribution of galaxies
along cosmic epochs, i.e. the so-called Large Scale Structure (LSS) of
the Universe, provides one of the most powerful constraints in
observational cosmology. Galaxy surveys \citep[see,
  e.g.,][]{Sanchez06} have also shown the relevance of galaxy
correlation functions, and in particular the two-point correlation
function (2pCF), as a useful statistical estimator to characterize the
LSS. The redshift evolution of galaxy 2pCF seems to be weaker/slower
that the expected evolution of the clustering for the underlying mass
distribution in concordance cosmologies \citep{Jenkins98}: this
mismatch is usually interpreted as the effect of physical mechanisms
acting on the baryonic component. This implies that the redshift
evolution of the 2pCF can also provide important constraints for
baryon physics and galaxy formation models. Indeed, successful
theoretical models of galaxy evolution on cosmological volumes usually
aim to reproduce the redshift evolution of key statistical estimators,
which are revealing of the correct build up of the stellar mass
(M$\star$) as a function of redshift, like the galaxy stellar mass
function (GSMFs) or luminosity functions (LFs). Nonetheless, it is
also of fundamental importance that these models are able to correctly
describe how the different galaxy types populated host dark matter
(DM) haloes (and sub-haloes) in different density environments
\citep[see, e.g.,][]{ColeKaiser89}.

According to the classical interpretation of the 2pCF shape, it can be
seen as a combination of \citep[see, e.g.,][]{CooraySheth02}: a 2-halo
term - corresponding to pairs of galaxies belonging to independent
halos - and a 1-halo term - coming from pairs of galaxies living in
the same halo. The 2-halo term is dominant at large scales
(separations larger than $\sim$3 Mpc) and comes mainly from pairs of
central galaxies. The 1-halo term is responsible for the 2pCF shape at
$\lesssim$2 Mpc and it is statistically dominated by galaxy pairs
belonging to the same halo (that is at least one of the two galaxies
is a satellite). 
Observationally, only the projected 2pCF - $w(r_{\rm p})$ - is
directly accessible, while the tridimensional 2pCF can be recovered
via de-projection. Early results by \citet{DavisGeller76} already
showed a clear dependence of clustering amplitude with galaxy type, by
showing that early type galaxies are more clustered than late-type
galaxies. The advent of large area redshift surveys like the Sloan
Digital Sky Survey (SDSS) provided the statistics to explore the
dependence of $w(r_{\rm p})$ amplitude and shape on galaxy properties
such as luminosity \citep[see, e.g.,][]{Zehavi02} or stellar mass
\citep{Li06}, showing that brighter/more massive galaxies are more
clustered than their fainter/less massive counterparts. A stronger
difference in the clustering signal is found when considering galaxies
of different colour \citep[see, e.g.,][]{Li06}; at fixed M$_\star$ (or
luminosity), red galaxies sensibly more clustered than blue
galaxies. Galaxy colour is sensitive to the star formation rate (SFR)
of model galaxies, which is regulated by a complex network of internal
and external physical mechanisms. Therefore, clustering predictions
for the active and passive galaxy populations represent a additional
diagnostic to assess whether a given model reproduces the
observational data in terms of star formation quenching as a function
of the environment.

These results have been then confirmed and extended to higher
redshift, thanks to redshift surveys such as the Galaxy And Mass
Assembly (GAMA), VIMOS\footnote{Visible Multi Object Spectrograph}
Public Extragalactic Redshift Survey (VIPERS), the VIMOS-VLT Deep
Survey (VVDS), the VIMOS Ultra Deep Survey (VUDS), the Deep
Extragalactic Evolutionary Probe (DEEP2), the spectroscopic redshift
Cosmic Evolution Survey (zCOSMOS) and showing that these conclusions
generally hold up to z$\sim$1 \citep[see, among others,][]{Coil06,
  Pollo06, Meneux08, Meneux09, Marulli13, Skibba14, Farrow15, Coil17}
and providing the first estimates of the 2pCF at z$\gtrsim$2
\citep{Durkalec18}. Overall, there is still some debate about the
redshift evolution of the amplitude of the 2pCF: results from DEEP2
and VVDS favour a small evolution of clustering amplitude with
redshift, while VIPERS and zCOSMOS show negligible differences with
respect to z$\sim$0. It worth stressing the difficulties in comparing
clustering measurements at different redshift, due to the different
galaxy selection biases reflecting in the sampling of the different
galaxy population. Moreover, the difference in clustering amplitudes
of active and passive galaxies has been confirmed out to z$\sim$2:
\citet{Coil17} studied the clustering as a function of SFR, and they
found the same trends defined by colours at low-z, galaxies below the
star-forming main sequence are more clustered than galaxies above it
\citep[see also][]{Mostek13}.

In this paper, we exploit predictions from the latest version of the
GAlaxy Evolution and Assembly ({\sc gaea}), that couples an explicit
partitioning of the cold gas into its molecular and neutral phases
with an improved modelling of both cold gas accretion onto central
Super-Massive Black Holes (SMBHs), and gas stripping in dense
environments. We showed in our recent work \citep{DeLucia24, Xie24}
that this model realization correctly reproduces the observed
fractions of quenched galaxies up to z$\sim$3-4, while predicting
number densities of massive quiescent galaxies at z$\sim$3 that are
the largest among recently published models. We will thus take
advantage of the most recent determinations for the 2pCF at several
redshifts to assess if our model is also able to reproduce both the
global spatial distribution of galaxies in the LSS, and the
distribution of galaxy populations split by their star formation
activity. The latter test will represent a critical confirmation of
the activity of \gaea to correctly grasp the physical mechanisms
responsible for the regulation of star formation in galaxies.

Most of our previous \gaea realizations have been run on dark matter
merger trees extracted from the Millennium Simulation \citep[MS
  hereafter]{Springel05}. This realization is based on a cosmological
model whose parameters are offset with respect to more recent
measurements -- particularly for the normalisation of the density
fluctuations at the present day ($\sigma_8$) and $\Omega_{\rm m}$. In
this paper, we will introduce a novel \gaea realization, run on the
\pmill simulation \citep[\pms hereafter]{Baugh19}, featuring a bigger
cosmological volume, an updated cosmology, a better mass resolution
and a finer time sampling. In particular, we will discuss the redshift
evolution of key statistical quantities like the GSMFs or the AGN LFs
as well as of galaxy clustering, as a function of stellar mass and
star formation activity.

This paper is organized as follows. In Section~\ref{sec:simsam} we
will present the new \gaea realization run on merger trees extracted
from the \pms simulation. In Section~\ref{sec:conv} we compare these
predictions with previous realizations of the same model run on
different cosmological simulations. In Section~\ref{sec:clust} we
discuss the predicted clustering of model galaxies at different
redshift and as a function of different galaxy properties. Finally, we
present and discuss our conclusions in Section~\ref{sec:discconcl}.

\section{N-body simulations and the galaxy formation model}
\label{sec:simsam}
\begin{table*}
  \caption{Numerical and cosmological parameters for the simulations
    considered in this work: $\sigma_8$ the normalisation of the
    density fluctuations at the present day; $H_0$ the Hubble constant
    (in units [${\rm km \, s^{-1}} \, {\rm Mpc}^{-1}$]); $n_{\rm
      spec}$ is the spectral index of the primordial density
    fluctuations; $L_{\rm box}$ the simulation box length (in units
    [$(h^{-1} \, {\rm Mpc})$]); $N_{\rm p}$ the number of particles,
    $M_{\rm p}$ the particle mass (in units [$h^{-1}\, {\rm M_{\rm
          \odot}}$]).}
  \label{table:simparam}
  \renewcommand{\footnoterule}{} \centering
  \begin{tabular}{cccccccccc} 
    \hline
    & $\Omega_{\rm \lambda}$ & $\Omega_{\rm m}$ &  $\Omega_{\rm b}$ & $\sigma_{8}$ & $n_{\rm spec}$ & $H_0$ & $L_{\rm box}$ & $N_{\rm p}$ & Log($M_{\rm p}$)  \\
    \hline
    MS   & 0.75  & 0.25  & 0.0455  & 0.9    & 1.0    & 73    & 500     & $ 2160^{3} $ & $8.935$ \\
    MSII & 0.75  & 0.25  & 0.0455  & 0.9    & 1.0    & 73    & 100     & $ 2160^{3} $ & $6.838$ \\
    \pms & 0.693 & 0.307 & 0.04825 & 0.8288 & 0.9611 & 67.77 & 542.16  & $ 5040^{3} $ & $8.025$ \\
    \hline
  \end{tabular}
\end{table*}
\begin{table*}
  \caption{Parameter Calibration chart. The value of the parameters
    are identical in all runs presented in this paper.}
  \label{tab:params}
  \renewcommand{\footnoterule}{} \centering
  \begin{tabular}{ccc}
    \hline
    Parameter & Value & Meaning \\
    \hline
    $\alpha_{\rm SF}$ & 0.073 & Star Formation efficiency \\
    $\epsilon_{\rm reheat}$ & 0.28 & Reheating efficiency \\
    $\epsilon_{\rm eject}$ & 0.10 & Ejection efficiency \\
    $\gamma_{\rm reinc}$ & 0.99 & Reincorporation efficiency \\
    $\kappa_{\rm radio}/10^{-5}$ & 1.36 & Hot gas Black Hole accretion efficiency \\
    $f_{\rm lowJ}/10^{-3}$ & 3.18 & Cold gas angular momentum loss efficiency \\
    $f_{\rm BH}/10^{-3}$ & 0.11 & Black hole accretion rate from reservoir \\
    $\epsilon_{\rm qw}/10^{2}$ & 4.86 & Quasar wind efficiency \\
    $f_{\rm cen}/10^{-3}$ & 3.39 & fraction of ISM added to the BH reservoir
as a consequence of AGN-driven outflows \\
    rps$_{\rm time}$ & 444 & Timescale of hot gas ram-pressure stripping \\
    kesi$_{\rm slow}$ & 1.19 & Ratio between specific angular momentum of
gas cooling through ‘slow mode’ and that of the halo\\
    kesi$_{\rm rapid}$ & 1.38 & Ratio between specific angular momentum of
gas cooling through ‘rapid mode’ and that of the halo\\
    \hline
  \end{tabular}
\end{table*}

\subsection{Numerical Simulations}
In this work, we will focus on predictions from a \gaea realization
coupled with merger trees extracted from the \pms N-body simulation
\citep{Baugh19}. Moreover, we will compare our results against
predictions from two other \gaea runs, based on the {\sc millennium}
Simulation \citep[MS]{Springel05} and {\sc millennium ii} Simulation
\citep[MSII]{BoylanKolchin09}. These runs corresponds to the
predictions at the basis of our recent work \citep{DeLucia24,
  Xie24}. Details on the assumed cosmologies, box size and particle
resolution for each N-body simulation are listed in
Table~\ref{table:simparam}.

The \pms follows the evolution of the matter distribution in a
volume of 800$^3$ Mpc$^3$, employing roughly 128 billion particles to
trace the assembly of the Dark Matter Haloes (DMHs). In terms of mass
resolution ($m_p = 1.06 \times 10^8 \hmsun$), the \pms provides
intermediate results to the MS and MSII. Moreover, it also provides an
updated cosmology, based on the first year results from the Planck
satellite \citep{Planck_cosmpar}, with respect to the MS and MSII,
that were run assuming cosmological parameters consistent with
first year results of the WMAP satellite \citep{Spergel03}.

The initial conditions for the \pms have been generated at z=127 using
second order Lagrangian perturbation theory \citep[see][for more
  details]{Jenkins13}, and the simulation has been run using a reduced
memory version of the N-body code {\sc gadget} \citep{Springel05}.
Halo and Sub-Halo catalogues have been extracted on the fly using a
classical FOF algorithm and the substructure finder code {\sc subfind}
\citep{Springel01}: halos and sub-haloes information have been saved
on a regular grid of 271 output times (i.e. snapshots), equally spaced
in expansion factor. This has to be compared with the 64-68 snapshots
of the MS and MSII over the same redshift range. The gain in temporal
(redshift) resolution is particularly relevant at z$\gtrsim$4 where
the \pms has $\sim$100 snapshots against the $\sim25$ of the
MS/MSII. In order to run GAEA on the PMS volume, we have constructed
subhalo based merger trees using the same approach described in
\citet{Springel05}. We refer to this paper for a detailed description
of the algorithm. In summary, for any given subhalo, we identify a
unique descendant by finding all subhalos at the subsequent output
that contain its most bound particles. We then give higher weight to
those particles that are more tightly bound to the subhalo under
consideration, and select as descendant the subhalo containing a
larger fraction of the most bound particles. In this way, we aim at
tracing the fate of the inner regions of the substructure, that might
survive for longer time after it has been accreted onto a larger
structure. Following \citep{Springel05}, we search for descendants in
the two following outputs to deal with small subhalos that fluctuate
below and above the assumed detection threshold (20 bound particles).
Once a unique descendant has been identified for all subhalos in the
simulated volume and for all available outputs, links to all
progenitors are automatically established, and merger trees can be
built. We adopt the same merger tree structure defined for the MS
suite and shown in Fig. 11 of \citep{Springel05}. The merger trees
adopted in this work have been constructed using half of the available
outputs. While reducing the number of redshifts available, this choice
reduces the memory and computational requirements of our GAEA runs. In
a forthcoming paper (Cantarella et al., in preparation), we will show
that results are not significantly different from those obtained using
merger trees constructed considering all available snapshots.

\begin{figure*}
  \centerline{ \includegraphics[width=9cm]{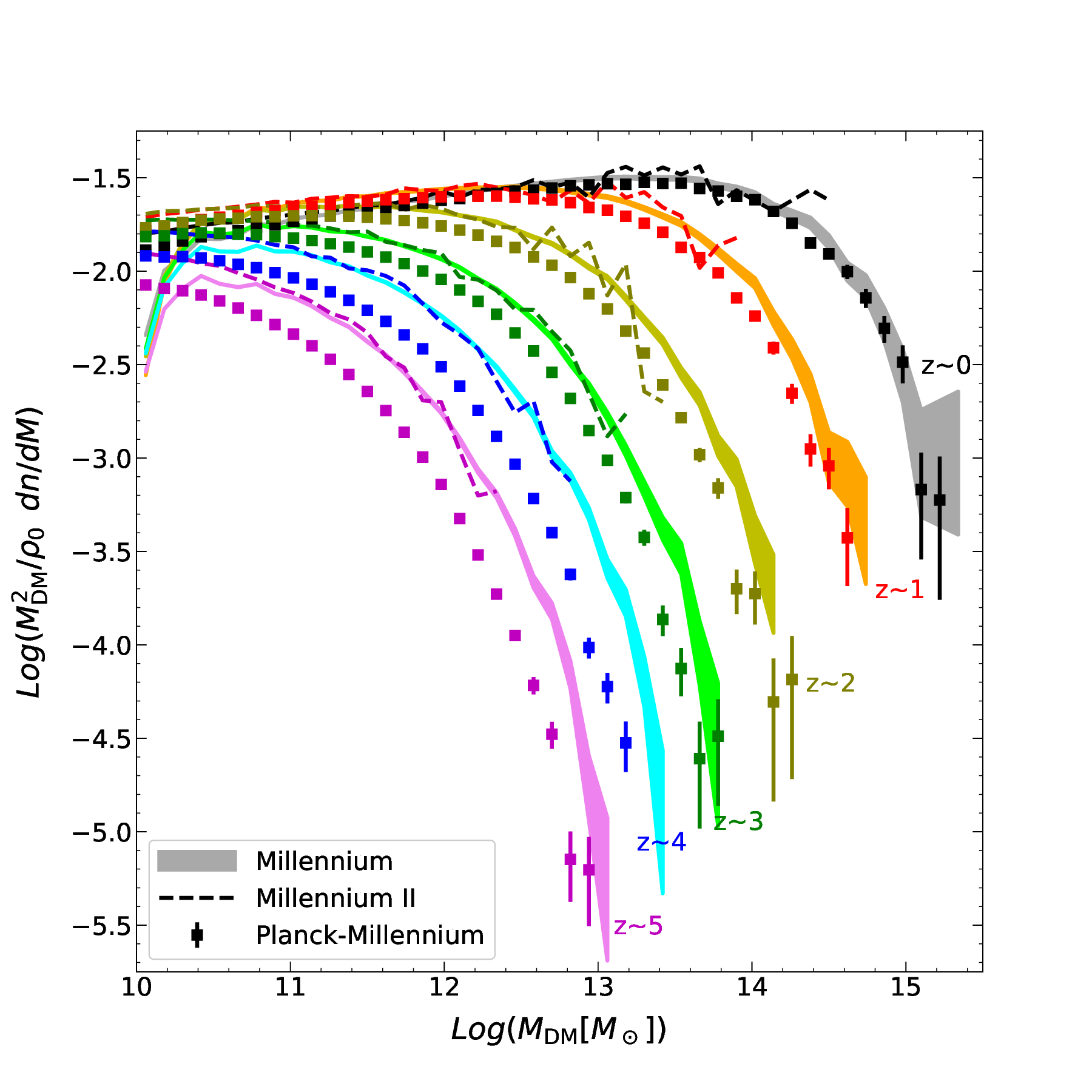}
    \includegraphics[width=9cm]{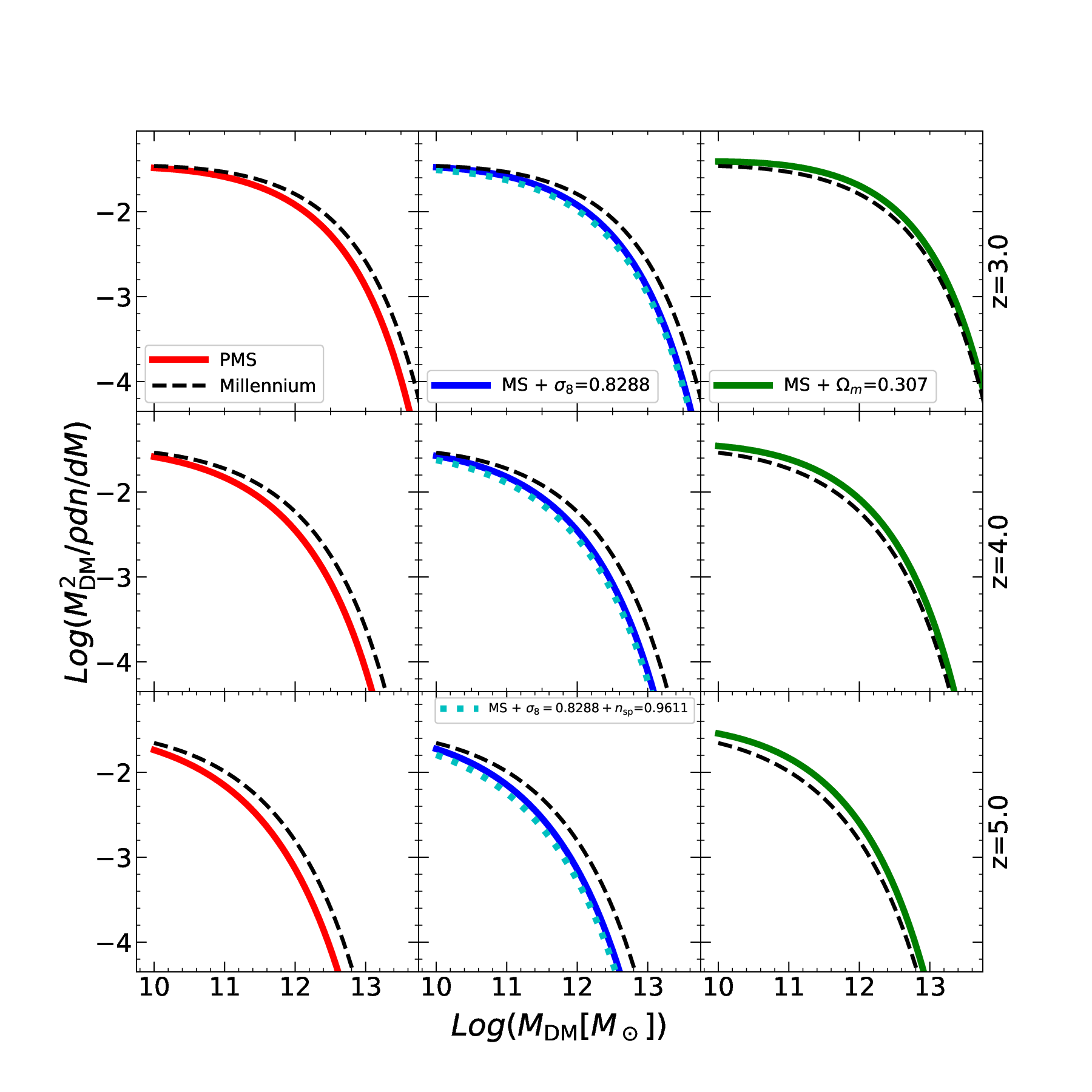} }
  \caption{{\it Left Figure:} Dark Matter HMF for the different
    simulations considered in this paper. Shaded regions show
    measurements based on the MS (1-$\sigma$ poissonian scatter);
    dashed lines correspond to the MSII simulation and squares with
    errorbars for the \pms simulation (1-$\sigma$ poissonian
    scatter). {\it Right Figure:} Impact of different cosmological
    parameters on the evolution of the HMF. The left-most panels show
    the HMFs expected for the MS and \pms cosmological models (dashed
    black and red solid lines respectively). In the central and
    right-most panels, the HMFs for the MS is compared with runs with
    theoretical expectations obtained modifying the following
    parameters: $\sigma_8$ (assumed to be the same as in the \pms;
    blue solid line); $\sigma_8$ and the spectral index (as in the
    \pms; cyan dotted line), $\Omega_{\rm m}$ (as in the \pms; green
    solid line).}\label{fig:dmh_mf}
\end{figure*}

In Fig.~\ref{fig:dmh_mf} (left panel), we compare the DM halo mass
functions (HMFs) derived from the {\sc subfind} subhalo group
catalogues for the MS (shaded areas), MSII (dashed lines) and \pms
(point with poissonian errorbars), in the redshift range 0<z<5. The MS
and MSII agree well over the mass range where both simulations have
enough volume and resolution. On the other hand, the \pms shows less
structures at high redshift and a faster evolution of the mass
function, that is very close to that computed from the MS at
z$\sim$0. In order to understand the origin of these differences we
consider in Fig.~\ref{fig:dmh_mf} (right panel) the evolution of the
HMF shape for varying cosmological parameters. We evaluate analytic
fits to the HMF at $3<z<5$ by assuming the \citet{Tinker08} model and
using the {\sc HMFcal} tool developed by \citet{Murray13}. In all
panels, the dashed black lines refer to HMFs predicted assuming the MS
cosmology. The red solid lines in the left panels correspond to HMFs
in the \pms cosmology. In the other panels, we show the HMFs predicted
by varying only a few selected parameters with respect to the MS
cosmology.  In particular we consider MS models with $\sigma_8=0.8288$
(blue solid lines), with $\sigma_8=0.8288$ and $n_{\rm spec}=0.9611$
(cyan dotted lines), with $\Omega_{\rm m}=0.307$ (green solid
lines). This comparison highlights that the differences between MS and
\pms are mainly driven by the lower value of $\sigma_8$ used in
\pms. The difference in $n_{\rm spec}$ and $\Omega_{\rm m}$ result in
(smaller) opposite effects, enhancing and contrasting the trends
induced by the different $\sigma_8$, respectively.

\subsection{\gaea}
\begin{figure*}
  \centerline{ \includegraphics[width=9cm]{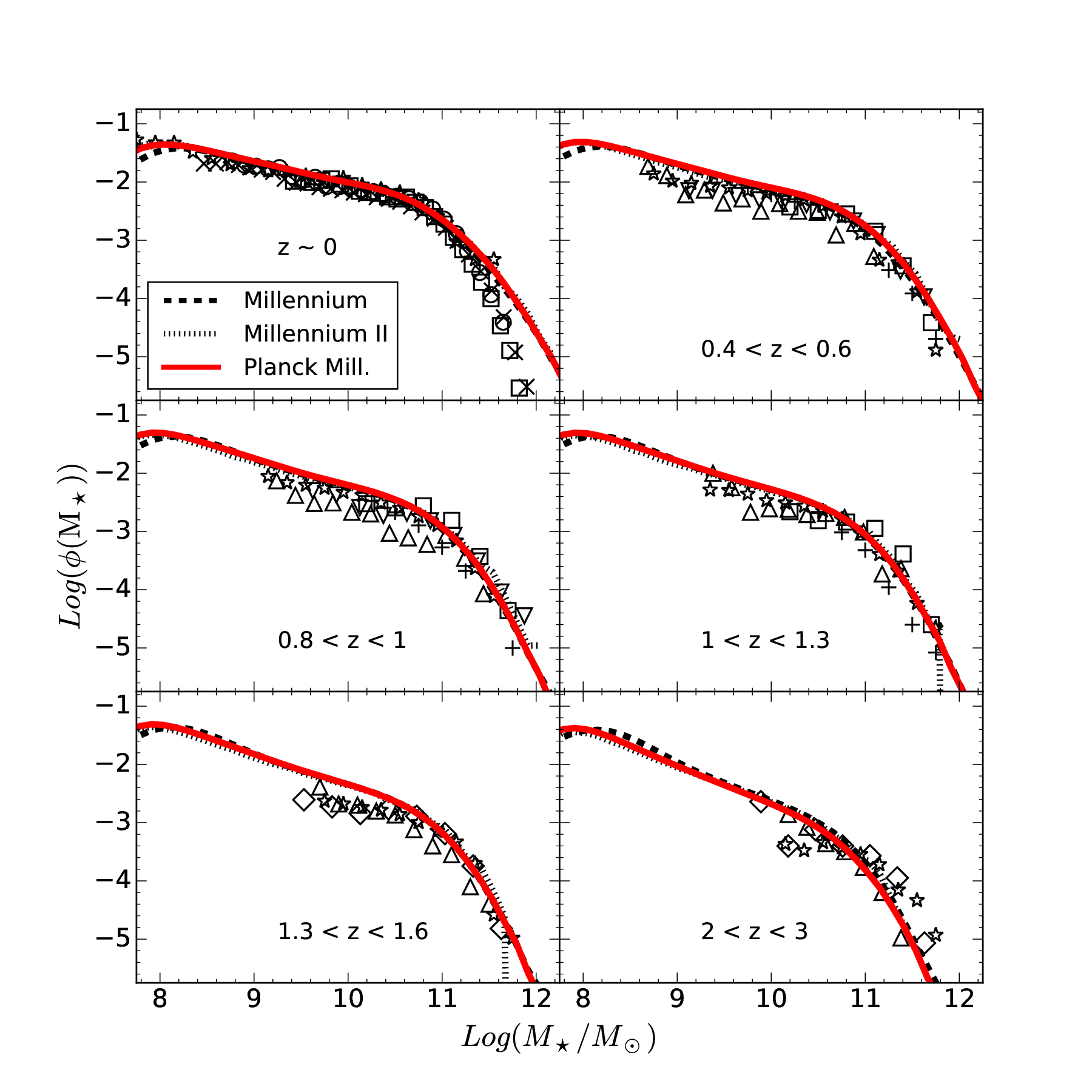}
    \includegraphics[width=9cm]{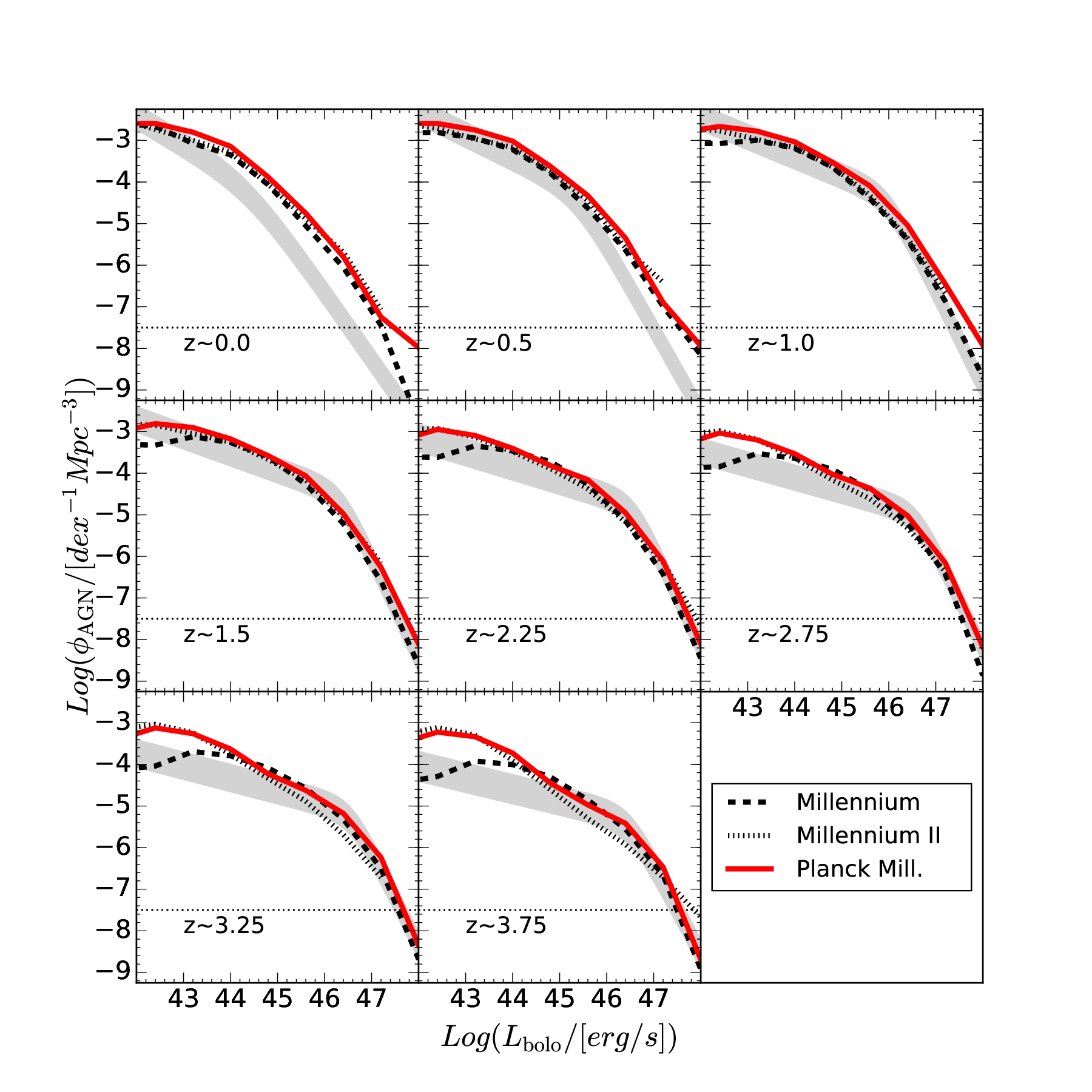} }
  \caption{{\it Left panel:} Redshift evolution of the galaxy stellar
    mass function; datapoints correspond to the compilation used in
    \citet[][see complete reference list for their
      Fig.~1]{Fontanot09b}. {\it Right panel:} Redshift evolution of
    the bolometric QSO/AGN-LF; the shaded regions show the expected
    space density from the empirical estimate of \citet{Shen20}. Black
    solid, black dashed and red solid lines refer the \gaea model
    version presented in DL24 and run on the MS, MSII and \pms
    simulations respectively.}\label{fig:allconvA}
\end{figure*}

In this study we consider predictions from our latest rendition of the
\gaea semi-analytic model (SAM). The approach starts from a
statistical description of the assembly of the Large Scale Structure
(i.e. a merger tree) that can be either analytically derived or
extracted from numerical experiments. Galaxies are assumed to form
within DM haloes, under the effect of a complex network of physical
processes responsible for energy and mass exchanges among different
baryonic components. SAMs follow the evolution of galaxy populations
by numerically solving a system of differential equations, aiming at
describing specific physical mechanisms, using parametrizations
derived from empirical, numerical or theoretical arguments. The
parameters involved are calibrated against a well defined set of
observational constraints, by sampling efficiently the parameter
space. The main advantage of the SAM approach lies in its reduced
computational demand, with respect to an hydro-dynamical simulations:
they are an optimal tool for studying galaxy evolution over large
cosmological volumes, with enough flexibility to extensively test
different physical models, explore their associated parameter space,
and therefore the impact of individual physical mechanisms to set the
observed properties of different galaxy populations.

\gaea builds from the model originally presented in
\citet{DeLuciaBlaizot07}, but now includes several improvements and
further developments. In particular: (a) the modelling of chemical
enrichment, based on a non-instantaneous treatment of the ejection of
gas, metals and energy that accounts for the mass dependence of
stellar lifetimes \citep{DeLucia14}; (b) an improved treatment for
stellar feedback, partially based on results of high-resolution
hydrodynamical simulations \citep{Hirschmann16}; (c) a detailed
tracing of the evolution of angular momentum \citep{Xie17}. The latest
rendition of the model has been presented in \citet[][DL24
  hereafter]{DeLucia24} and merges independently developed versions,
namely (d) the \citet{Xie17} explicit partition of the cold gas into
its atomic and molecular components; (e) the \citet{Xie20} treatment
for the non-instantaneous stripping of cold and hot gas in satellites
galaxies; and (f) the \citet{Fontanot20} modelling for cold gas
accretion onto super-massive Black Holes (SMBH) and the onset of
AGN-driven outflows. This new version of the \gaea model has been
calibrated on the $z<3$ evolution of the galaxy stellar mass function
(GSMF), on the $z<4$ evolution of the AGN luminosity function (LFs)
and on the local HI and H2 mass function. DL24 shows that it correctly
reproduces the evolution of the fraction and number densities of
quenched galaxies up to $z\sim4$, providing the largest number
densities of massive galaxies at $z>3$ among similar theoretical
models, predicting the rise of the first massive quenched galaxies at
$z\sim6-7$ \citep{Xie24}. A more throughout analysis of the properties
of the highest-z galaxies, especially in terms of their predicted UV
luminosities as seen by JWST, will be presented in forthcoming work
(Cantarella et al., in preparation). We stress that the latest \gaea
rendition preserves the good agreement with a variety of observational
constraints that have been explored in recent years, such as the
evolution of the mass-metallicity relations \citep[][both for the
  gaseous and stellar phases - see also
  Fig.~\ref{fig:allconvB}]{Fontanot21}, the $z>3$ GSMF and the cosmic
star formation rate density up to the highest redshifts for which
observations are available \citep{Fontanot17b}.

\section{Convergence of model predictions on different simulations}
\label{sec:conv}
\begin{figure*}
  \centerline{ \includegraphics[width=9cm]{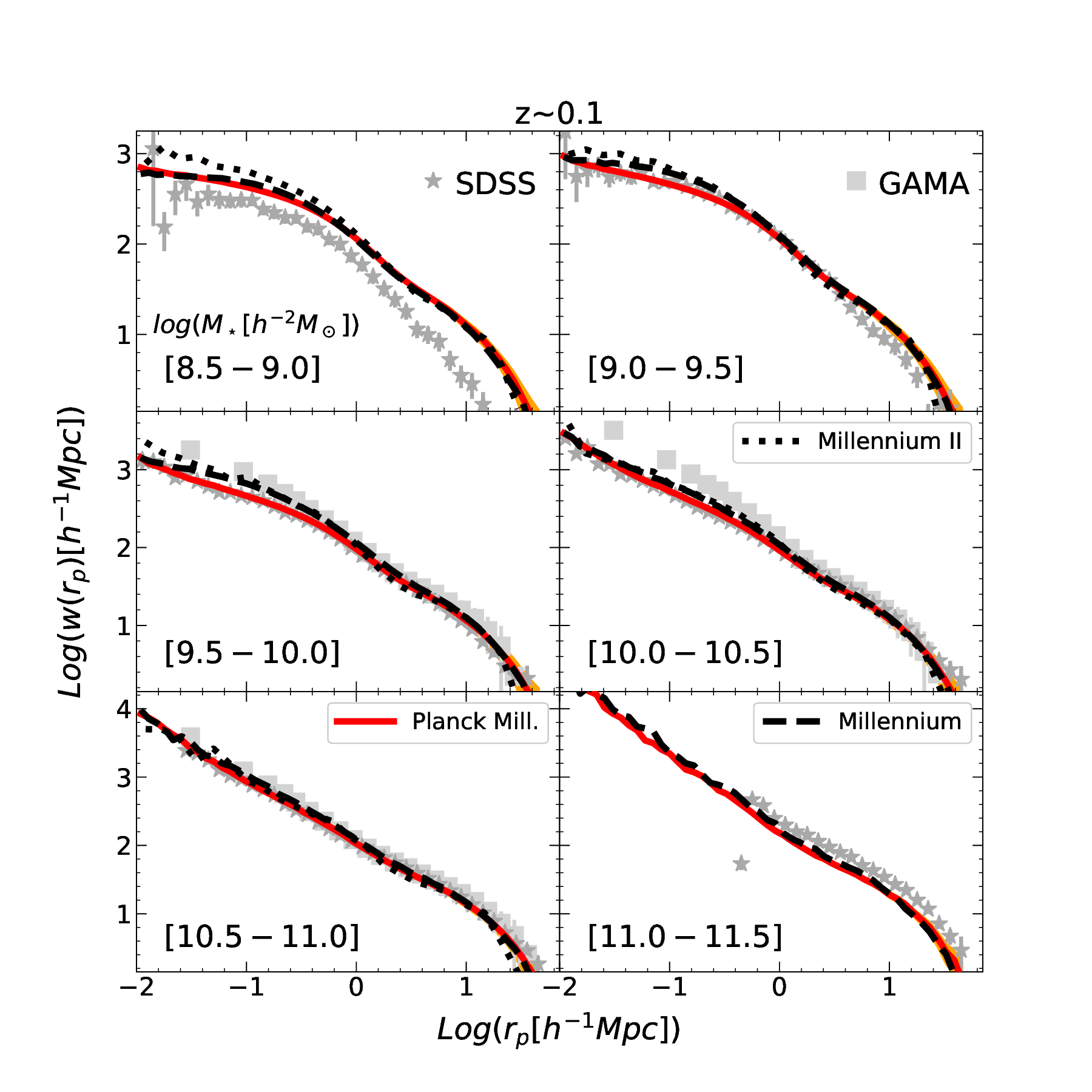}
  \includegraphics[width=9cm]{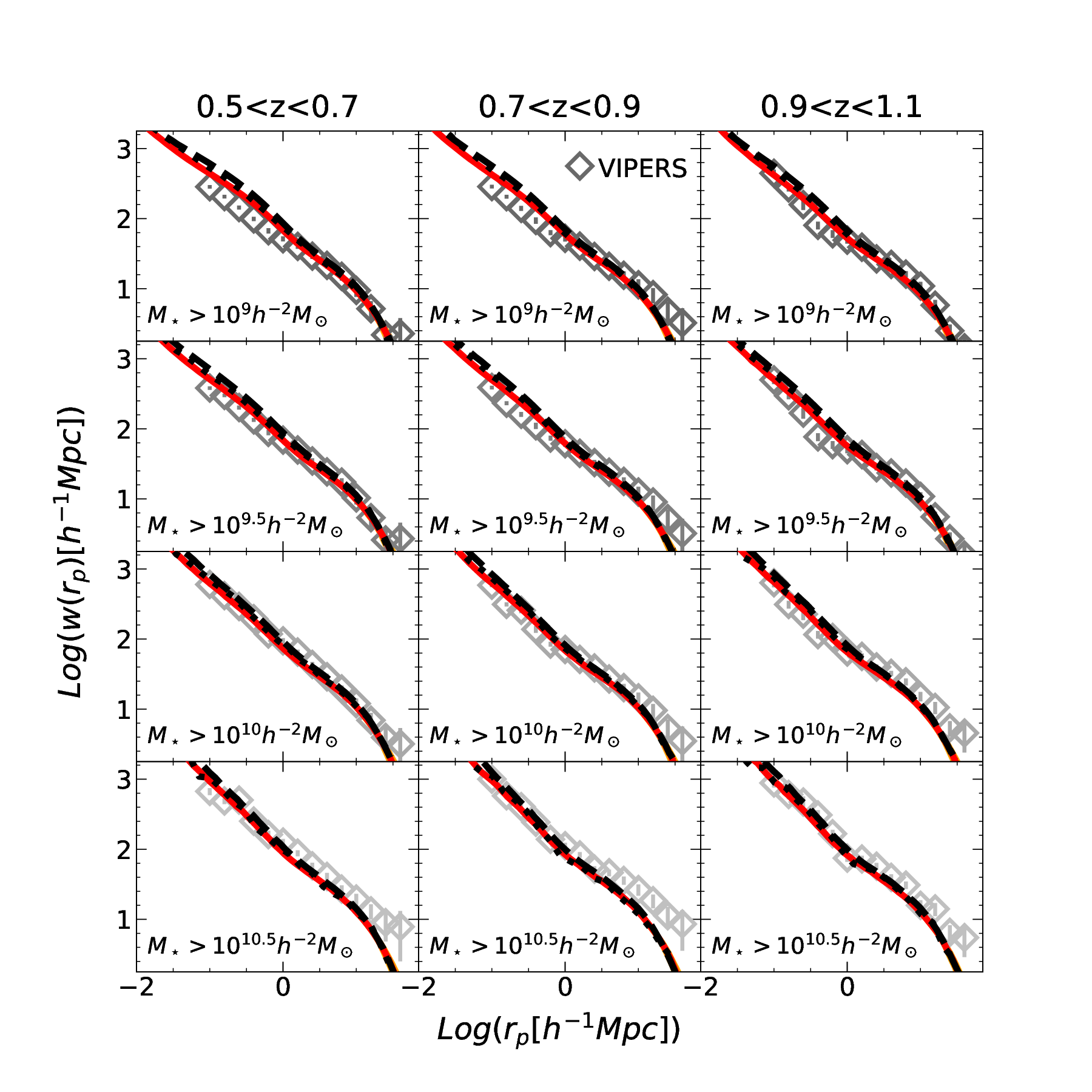} }
  \centerline{ \includegraphics[width=9cm]{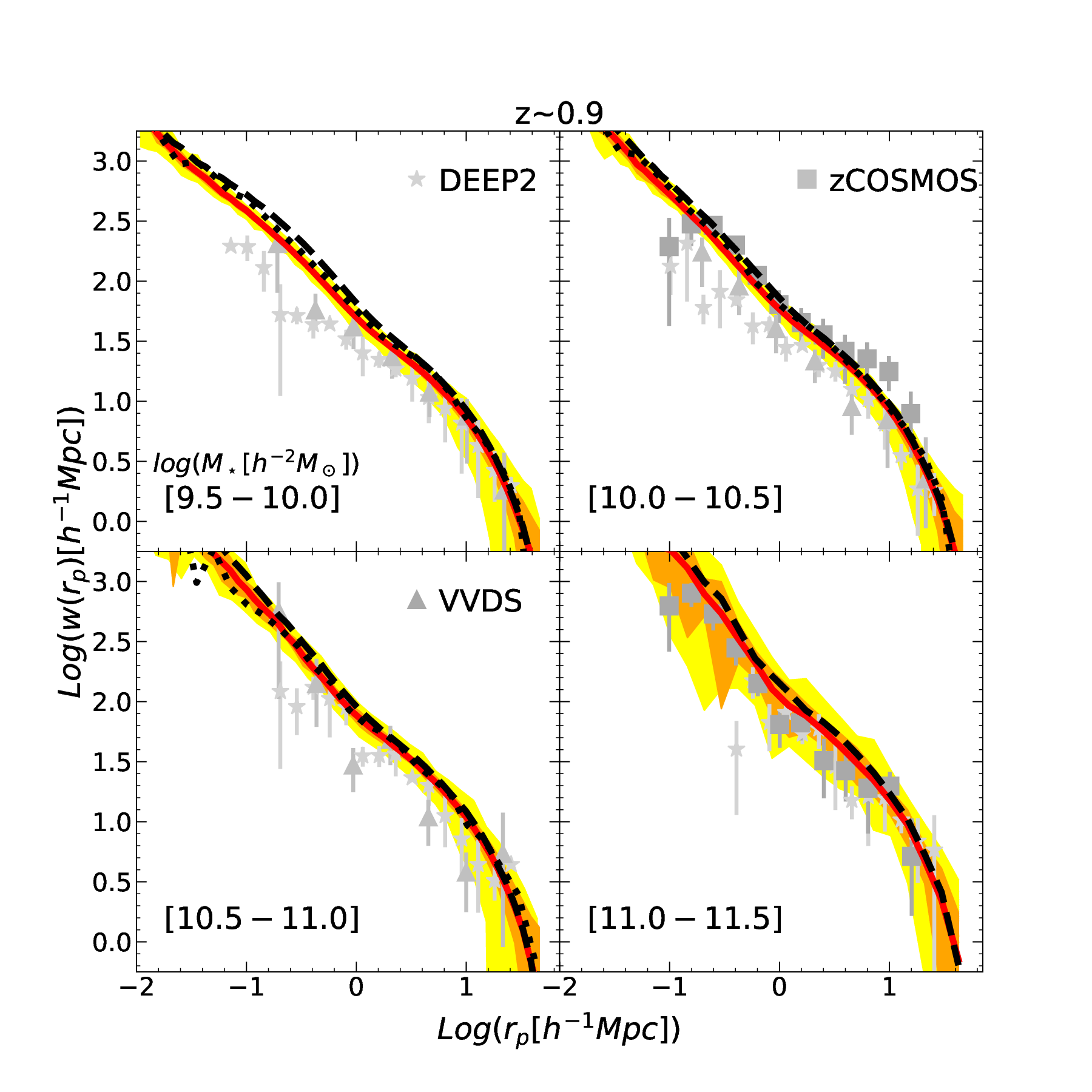}
    \includegraphics[width=9cm]{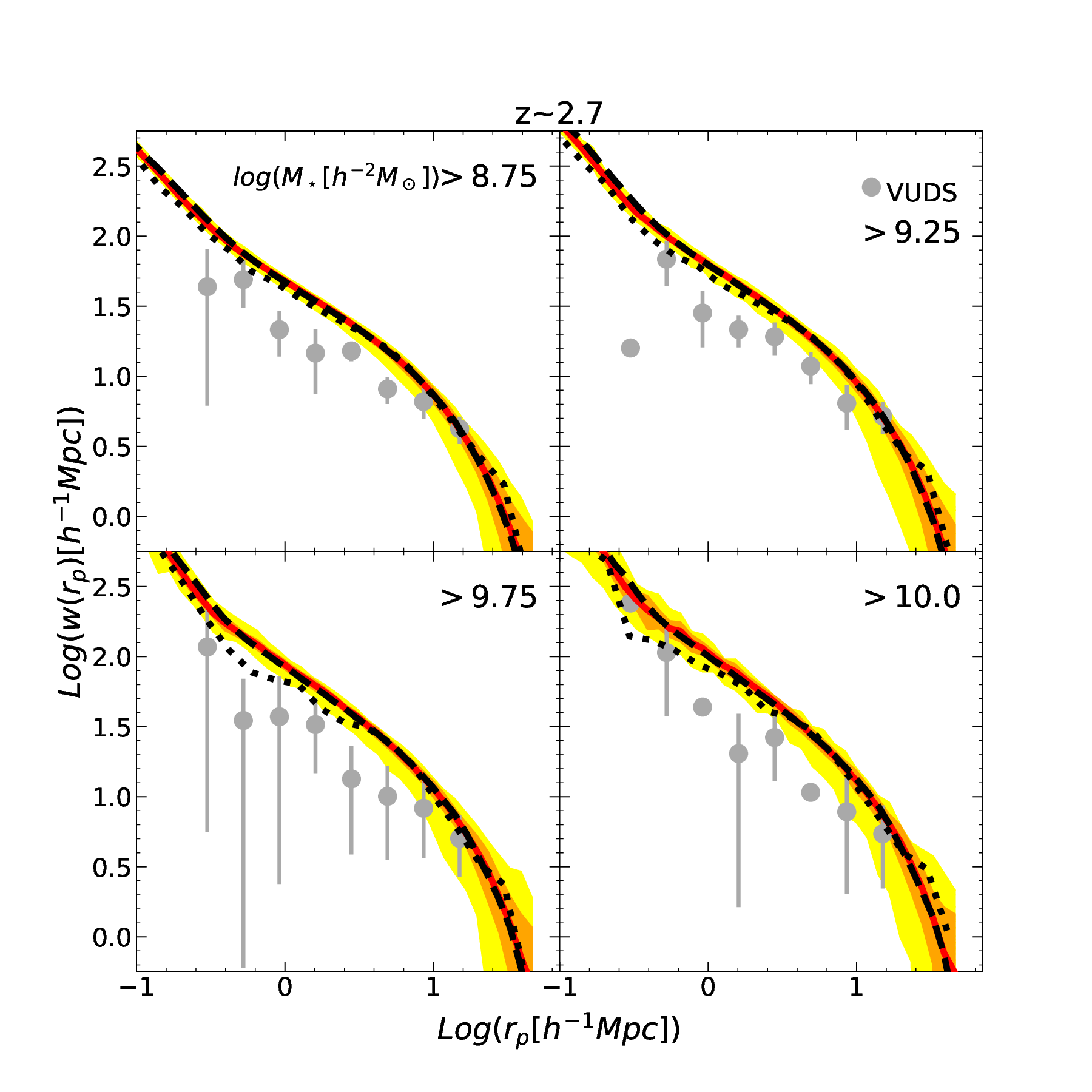} }
  \caption{Redshift evolution of the projected 2-point correlation
    function at different bins of stellar mass (as labelled). {\it
      Upper left panel}: data from SDSS at z$\sim$0 \citep{Li06}. {\it
      Upper right panel}: data from VIPERS at 0.5<z<1.1
    \citep{Marulli13}. {\it Lower left panel}: data at z$\sim$0.9 from
    VVDS \citep{Meneux08}, zCOSMOS \citep{Meneux09}, DEEP2 obtained
    using the same techniques as in \citet{Li06}. {\it Lower right
      panel}: data from VUDS at 2<z<3.5 \citep{Durkalec18}. \gaea
    predictions refer to the realizations run on the MS (black dashed
    line) and on the \pms (red solid line) as in
    Fig.~\ref{fig:allconvA}. The blue dot-dashed line refers to the
    2-pCF computed using central galaxies belonging to DMHs in a mass
    range derived from the stellar mass range using the
    \citet{Behroozi13} relation (see main text for details). Orange
    and yellow areas correspond to the jackknife estimated errors and
    cosmic variance dispersion respectively. }\label{fig:wpgen}
\end{figure*}
In order to appreciate the effect of a change in the cosmological
parameters and resolution on the predictions of our model, we consider
\gaea runs over merger trees extracted from the \pms, MS and MSII. All
predictions we discuss in the following sections are based on model
realization using the same values for the relevant physical parameters
involved in the SAM definition. These parameters have been calibrated
on the MS trees and correspond to the model presented in DL24. We list
the most relevant in Table~\ref{tab:params}. In the following, we will
show model predictions convolved with an estimated error on stellar
and gaseous masses of 0.25 dex, thereby taking the Eddington bias
\citep{Jeffreys38, Eddington40} into account. The uncertainty
considered represents a conservative choice especially at high-z. We
contrast model predictions from the different simulations with
compilations of observational data used in our previous works. In this
section (Fig.~\ref{fig:allconvA}), we focus on a basic set of physical
properties, whose evolution is crucial for understanding the process
of galaxy assembly, like the Galaxy Stellar Mass Function (GSMF - left
panel) from z$\sim$0 to z$\sim$3 and the AGN bolometric Luminosity
Function (LF, right panel) from z$\sim$0 to z$\sim$4. Those are among
the main observables used to calibrate our model. In Appendix~A, we
include additional observational constraints considered for model
calibration, plus several predictions we consider in our recent work.

It is remarkable that predictions corresponding to realizations run on
different simulations show such a high level of consistency. This
shows that our latest rendition of the \gaea model exhibits an
improved convergence with respect to previous versions. This is
remarkable for the predicted evolution of the AGN-LF: the space
density of faint AGNs is very sensible to the number of trigger events
(i.e. mergers) and in the past this has limited the convergence of our
model to relatively bright luminosities (see e.g. Fig.~A2 in
\citealt{Fontanot20}). The latest \gaea version shows a good
convergence between the MS and the \pms runs down to L$_{\rm bolo}
\sim 10^{42}$ erg/s at z<1, and down to L$_{\rm bolo} \sim 10^{44}$
erg/s at higher redshifts; predictions from the \pms and MSII agree
even better around the knee of the LF, while at brighter luminosities,
the MSII lacks volume to correctly reproduce the rarer sources. For all
other statistical estimators the level of convergence between the
different runs is high and in most cases predictions based on
different simulations are virtually indistinguishable. We interpret
this effect as a result of the interplay between the improved
modelling of star formation, based on the partition of the cold gas as
in \citet{Xie17}, and the new prescription for cold gas accretion onto
the central SMBHs \citep{Fontanot20}. Both physical mechanisms act as
regulators of the total amount of cold gas available for star
formation, chemical enrichment and SMBH accretion. In fact, versions of
the model implementing just one of the two physical mechanisms (and
run on different simulation) do not achieve the same level of
convergence over such a wide range of redshift and physical properties, as
that presented here.

\section{Clustering Analysis}
\label{sec:clust}
\begin{figure*}
  \centerline{ \includegraphics[width=9cm]{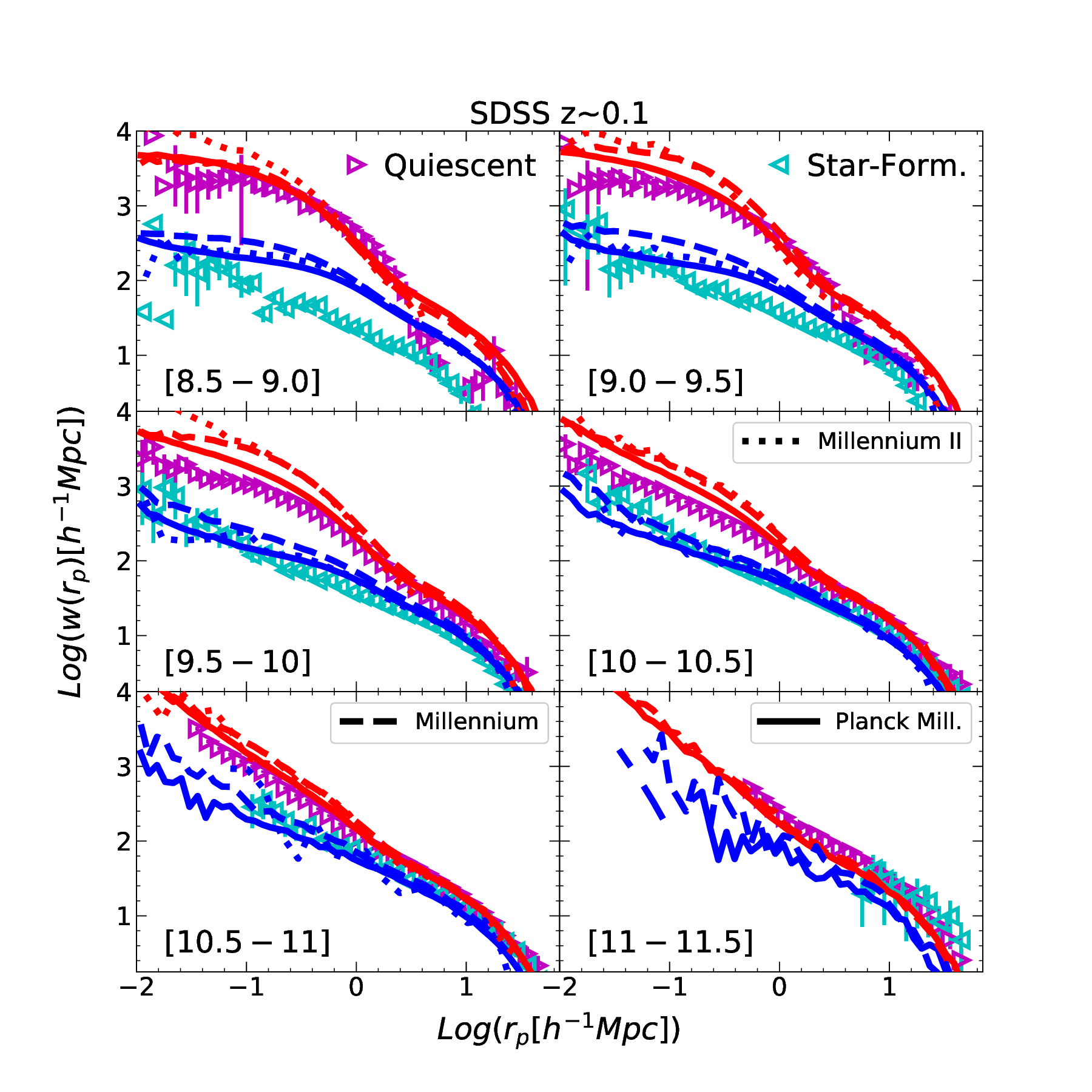}
    \includegraphics[width=9cm]{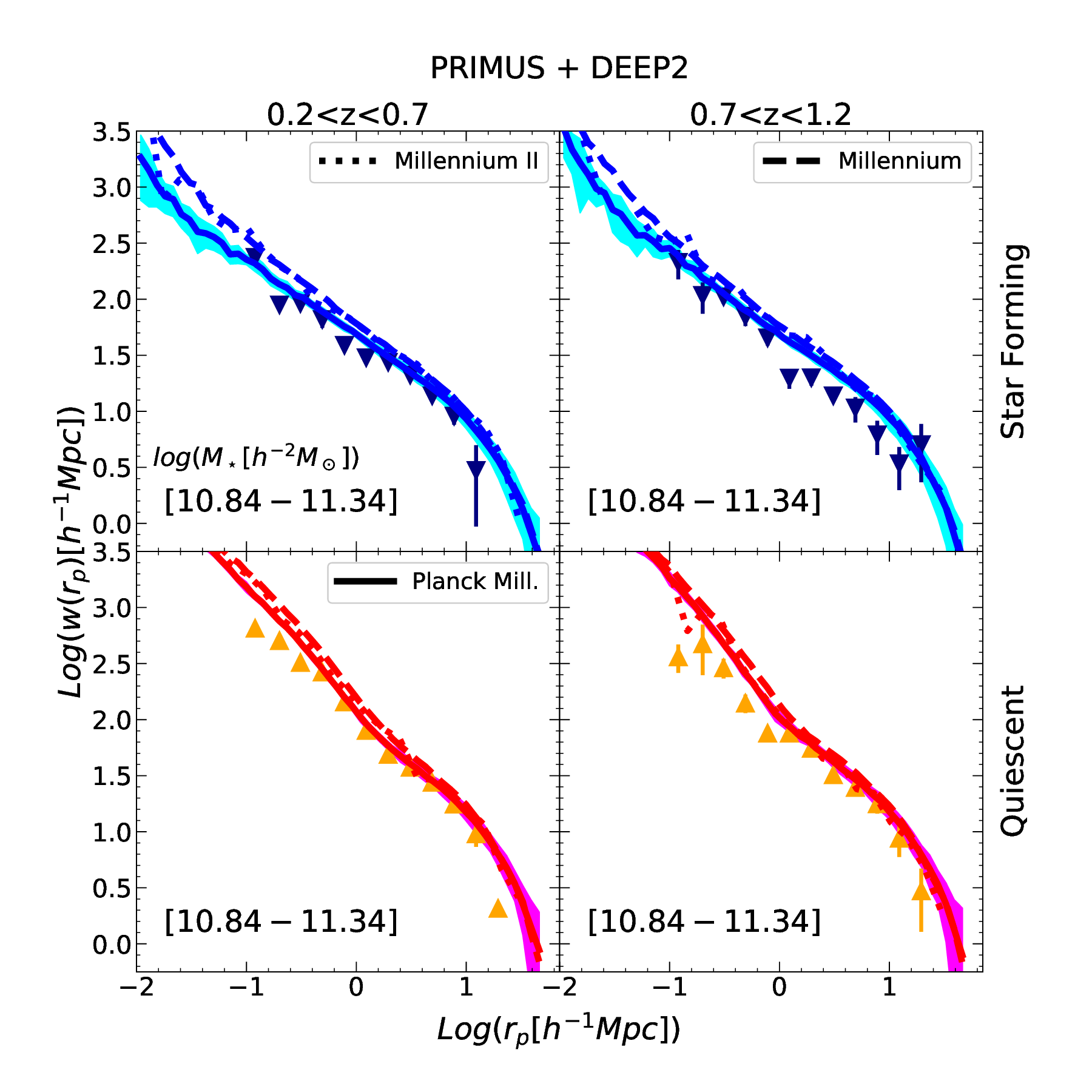} }
  \caption{{\it Left panel} Projected 2-point correlation function for
    red/blue galaxies at z$\sim$0.1; data from \citet{Li06}. {\it
      Right panel} Projected 2-point correlation function for star
    forming/quiescent galaxies; data from \citet{Coil17}. \gaea
    predictions refer to the realizations run on the MS (dashed lines)
    and on the \pms (solid lines) as in
    Fig.~\ref{fig:allconvA}.}\label{fig:wpcol}
\end{figure*}

\subsection{Redshift evolution of the 2-point correlation function}\label{sec:clust1}

We then consider the predicted correlation functions for our model
universes at different redshifts. In the following, we will focus on
the projected 2-point correlation function $w(r_p)$; this choice makes
the comparison with observational data straightforward, and does not
have to rely on de-projection algorithms. We study the evolution of
the projected correlation function using estimates from several
surveys: going from low-z to higher redshifts we compare against
results from SDSS \citep[z$\sim$0.1, maximum integration
  scale\footnote{The maximum integration scale $\pi_{\rm max}$
  represents the finite upper limit used for the integration of the
  real-space correlation function.} $\pi_{\rm max}$=40
  $\hmsun$]{Li06}, GAMA \citep[z<0.2, $\pi_{\rm max}$=40
  $\hmsun$]{Farrow15}, VIPERS \citep[$0.5<z<1.1$, $\pi_{\rm max}$=30
  $\hmsun$]{Marulli13}, DEEP2 \citep[z$\sim$0.9, $\pi_{\rm max}$=40
  $\hmsun$]{Li06}, VVDS \citep[z$\sim$0.9, $\pi_{\rm max}$=20
  $\hmsun$]{Meneux08}, zCOSMOS \citep[zCOSMOS][z$\sim$0.9, $\pi_{\rm
    max}$=20 $\hmsun$]{Meneux09} and VUDS \citep[z$\sim$2.7, $\pi_{\rm
    max}$=20 $\hmsun$]{Durkalec18}. We collect all these observational
measurements in Fig.~\ref{fig:wpgen} and we contrast them with the
prediction from \gaea realizations based on the MS (black dashed
lines), the MSII (dotted lines) or the \pms (red solid lines).  Where
not visible, the dotted lines (MSII) overlap with the dashed lines
(MSI). We do not show predictions from the MSII for bins corresponding
to M$_\star>$10$^{11} h^{-2}$ M$_\odot$, as there are too few galaxies
above this mass limit to provide a statistical estimate for the
2pCF. We compute $w(r_p)$ from \gaea using the python {\sc corrfunc}
package \citep{CorrFunc}; model galaxy samples have been extracted
from the simulated volumes considering all galaxies in the stellar
mass range indicated in the legend, i.e. we do not attempt to
replicate selection criteria specific of the different surveys
considered. In each plot (as well as in Fig.~\ref{fig:wpcol}
and~\ref{fig:wpgas}), model estimates have been computed using the
same $\pi_{\rm max}$ as in the observed samples we compare with, but
for the z$\sim$0.9 panel. In this case we show model predictions
corresponding to $\pi_{\rm max}$=40 $\hmsun$ (i.e. the value used in
the DEEP2 survey). In our analysis we include an estimate for the
typical uncertainty on M$_\star$ by convolving the intrinsic
predictions of the model with a log-normal error distribution of
amplitude 0.25 dex. We check that our main conclusions do not depend
on this choice: as $w(r_p)$ predictions based on intrinsic M$_\star$
provide a consistent picture. We also explore the impact of
uncertainties in the theoretical $w(r_p)$ determination by computing
the jackknife errors using the {\sc CosmoBolognaLib} package
\citep{CosmoBolognaLib}. We show these uncertainties for the \pms as
the orange areas. We also assess the impact of cosmic variance on the
comparison between data and model predictions: we randomly extract
from the \pms 100 boxes with the same volume as the observational
survey. The yellow areas in Fig.~\ref{fig:wpgen} highlight the range
covered by these sub-volumes. Overall, jackknife errors are relevant
at large separations and for the most massive galaxies. Cosmic
variance is relevant at the highest redshift considered and for small
surveyed volumes, where it leads to errors that are slightly larger
than those based on jackknife at intermediate scales.

Overall, the level of agreement between theoretical and observed
correlation functions is satisfactory over the entire redshift range
considered. Models run on the MS, MSII and \pms agree well with each
other, but for a slight tendency of the MS and MSII runs to predict
systematically stronger correlations than \pms. The effect is more
pronounced at small stellar masses and close separations, and we
ascribe it to the larger value of $\sigma_8$ used in the MS and
MSII. At z$<0.2$ (Fig.~\ref{fig:wpgen}, upper left panel), theoretical
$w(r_p)$ are very close to SDSS constraints for $10^9 \msun < {\rm
  M}_\star 10^{11} \msun$. Above and below this mass range, model
predictions slightly underestimate and over-estimate the measured
clustering signal, respectively.

The lower normalization of the 2pCF for galaxies more massive than
10$^{11} \msun$ is connected with the overprediction of the high-mass
end of the z$\sim$0 GSMF (top left panel of
Fig.~\ref{fig:allconvA}). In this mass range, our model galaxies lie
above the M$_{\rm vir}$ versus M$_\star$ relation that can be inferred
from data using a classical subhalo abundance matching technique
\citep{Behroozi13,Moster13}, i.e. \gaea massive galaxies tend to live
in DMHs less massive than what can be inferred using halo occupation
distribution statistics.

For galaxies in the lowest mass bin considered, the overprediction of
the clustering amplitude is mainly driven by the mass distribution of
satellites and centrals in the smallest resolved DM haloes, an effect
already discussed in \citet{Wang13b}. These authors showed that to
reproduce at the same time the GSMF and 2pCF in the framework of an
halo occupation distribution model, satellite galaxies should be less
massive than centrals, on average, at fixed stellar mass. They also
considered results from two independent semi-analytic models that
struggled to reproduce these results - a problem that is shared also
by our GAEA model.  It is worth stressing that \gaea lacks an explicit
treatment for tidal stripping of stars from satellite galaxies, a
process that could move model predictions in the right direction and
whose impact we plan to study in future work.  Moreover, at scales
$\lesssim$2 Mpc orphan galaxies (i.e. satellite galaxies whose
substructure has fallen below the resolution limit of the simulation)
represent a relevant population and their orbital evolution and
merging times estimates might be incorrect (see
e.g. \citealt{DeLucia10} for a discussion).

Similar conclusions hold at intermediate redshifts ($0.5<z<0.9$) for
VIPERS (Fig.~\ref{fig:wpgen}, upper right panel). The situation is
slightly more complex at z$\sim$0.9, where different surveys suggest
slightly different normalization for $w(r_p)$. This could be due to
the different surveyed volume or to the presence of overdensities
(like in the case of the zCOSMOS sample,
\citep{Meneux09}). Nonetheless, \gaea predictions lie within the
clustering measurements obtained from different surveys at z$\sim$0.9
(Fig.~\ref{fig:wpgen}, lower left panel) and the agreement with data
is acceptable to the smallest separations measured. The predicted
evolution of the $w(r_p)$ normalization at z$\lesssim$1 is rather
small: in most cases $w(r_p)$ can be well approximated by a single
power law over a wide separation scale; \gaea also captures the
observed flattening of the relation at low-z for the smallest
separations probed by the SDSS.

Moving to higher redshifts ($2<$z$<3.5$), a larger discrepancy of the
$w(r_p$) normalization can be appreciated with respect to the VUDS
data (Fig.~\ref{fig:wpgen}, lower right panel): \gaea predictions are
still consistent with observational measurements within the
uncertainties, but tend to lie systematically above the data. Cosmic
variance can also have a non-negligible impact here (the VUDS volume
correspond to $\sim$1.75 $\times$ 10$^7$ Mpc$^3$ roughly 1/175th of
the \pms). Overall, we can conclude that \gaea is able to satisfactory
reproduce the observed evolutionary trends of the correlation strength
up to z$\lesssim$3.5. This represents a non-trivial success for our
latest \gaea model release.

\subsection{Colour/SFR dependence of the 2-point correlation function}

As we already discussed, several authors (see e.g. \citealt{Wang13a})
pointed out that the clustering strength depends not only on
$M_\star$, but also on the star formation activity of galaxy
population, with red and quiescent galaxies being more clustered than
their blue and star forming counterparts with similar $M_\star$. In
Fig.~\ref{fig:wpcol}, we compare model predictions with SDSS data at
z$\sim$0.1 \citep[$\pi_{\rm max}$=40 $\hmsun$]{Li06} and data up to
z$\sim$1.0 \citep[$\pi_{\rm max}$=40 $\hmsun$]{Coil17} obtained
combining the PRIsm MUlti-object Survey \citep[PRIMUS][]{Coil11} with
DEEP2. At all redshifts, we consider model predictions corresponding
to quiescent and star-forming subsamples, using separation thresholds
based on their specific SFR (sSFR = SFR/M$_\star$), for consistency
with our previous work (DL24).

At z$\sim$0.1, model galaxies are classified using a sSFR= 10$^{-11}$
yrs$^{-1}$ separation. This is formally different than the choice for
SDSS galaxies, which have been split according to their g-r colour:
however, we checked that using a classifications based on galaxy
colours (e.g. as in \citealt{Springel18}) does not qualitatively
change our main results. Overall, \gaea realizations show a stronger
clustering signal of red galaxies with respect to blue ones, in
qualitative agreement with observations. In more detail, our models
reproduce well the clustering strength for red galaxies over the
entire mass range considered. This is particularly true for the \pms
realization, while some residual overestimate of the clustering signal
holds at the lowest masses, and smallest scales. It is worth stressing
that this is the range of scales and stellar masses where our model
samples are dominated by orphan galaxies (that are typically
associated with low levels of SFR and red colors). The clustering
strength of blue galaxies is also reasonably well reproduced for
intermediate masses (i.e. in the 10$^{9.5}$ < M$_\star$/[$\hmsun$] <
10$^{11}$ range). Outside this range, the clustering of star-forming
galaxies shows the same trends discussed for the global population,
thus suggesting that this galaxy population is driving the tensions
between predictions and data reported in Sec.~\ref{sec:clust1} and
Fig.~\ref{fig:wpgen}. At higher redshift, we contrast clustering
estimates from the PRIMUS and DEEP2 surveys with \gaea predictions by
splitting model galaxies into a quiescent and star forming subsamples,
using the same (redshift-dependent) sSFR threshold employed in the
data \citep{Coil17}:

\begin{equation}
{\rm Log (SFR)} = - 1.29 + 0.65 \times ({\rm Log M}_\star - 10) + 1.33 \, (z - 0.1)
\end{equation}

\noindent
The agreement between data and model predictions in the single mass
bin available ($10.5<$Log(M$_\star [h^{-1} \msun$]) $<11$) is
very good, both for the red and blue populations.
\begin{figure}
  \includegraphics[width=9cm]{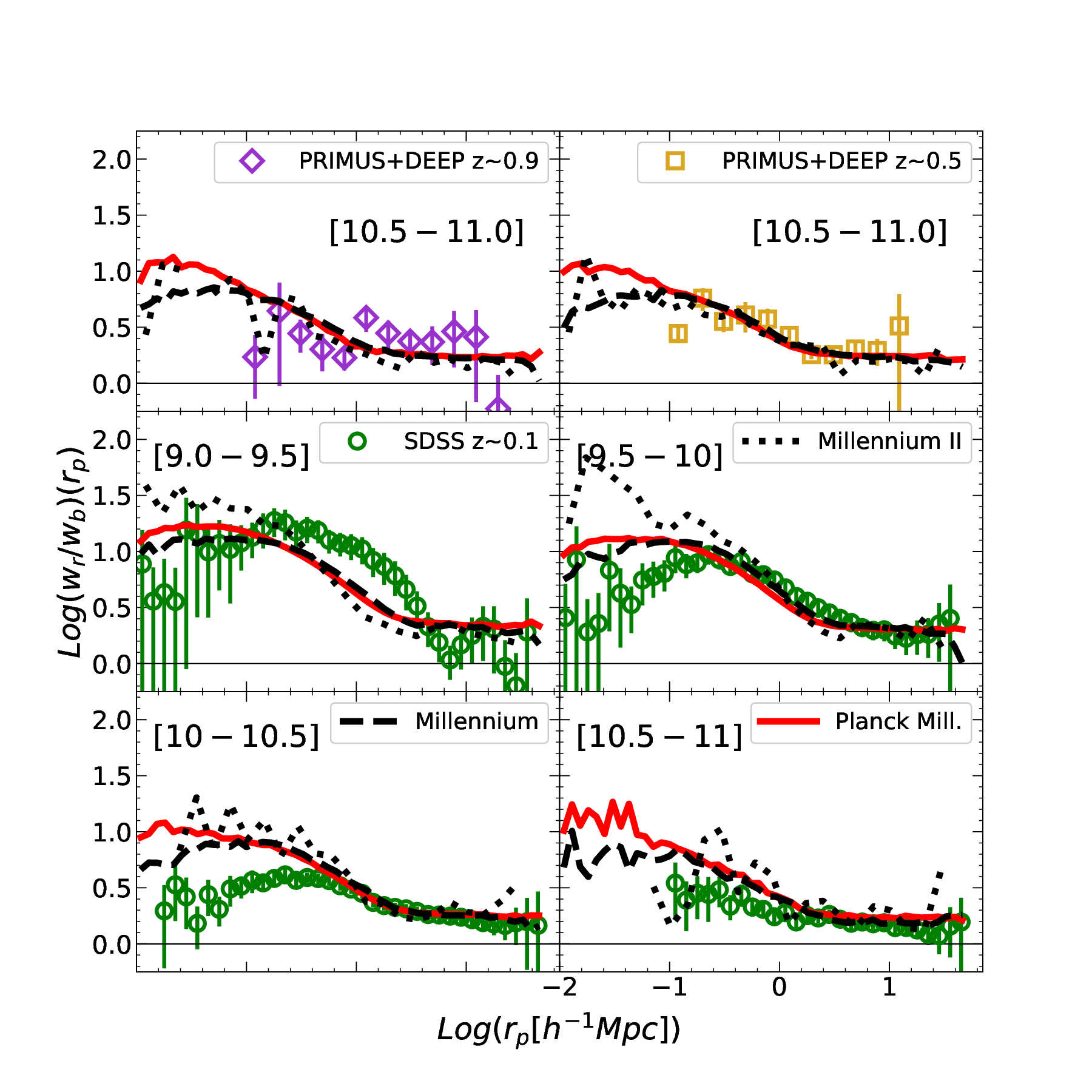}
  \caption{Ratio between the projected 2-point correlation function
    for red and blue galaxies. Datapoints have been arranged from
    Fig.~\ref{fig:wpcol}: green points refer to data from
    \citet[][SDSS at z$\sim$0.1]{Li06}; diamonds and squares to data
    from \citet[PRIMUS+DEEP sample at z$\sim$0.5 and
      z$\sim$0.9]{Coil17}. \gaea predictions refer to the realizations
    run on the MS (black dashed line) and on the \pms (red solid line)
    as in Fig.~\ref{fig:allconvA}.}\label{fig:wprat}
\end{figure}
\begin{figure*}
  \includegraphics[width=18cm]{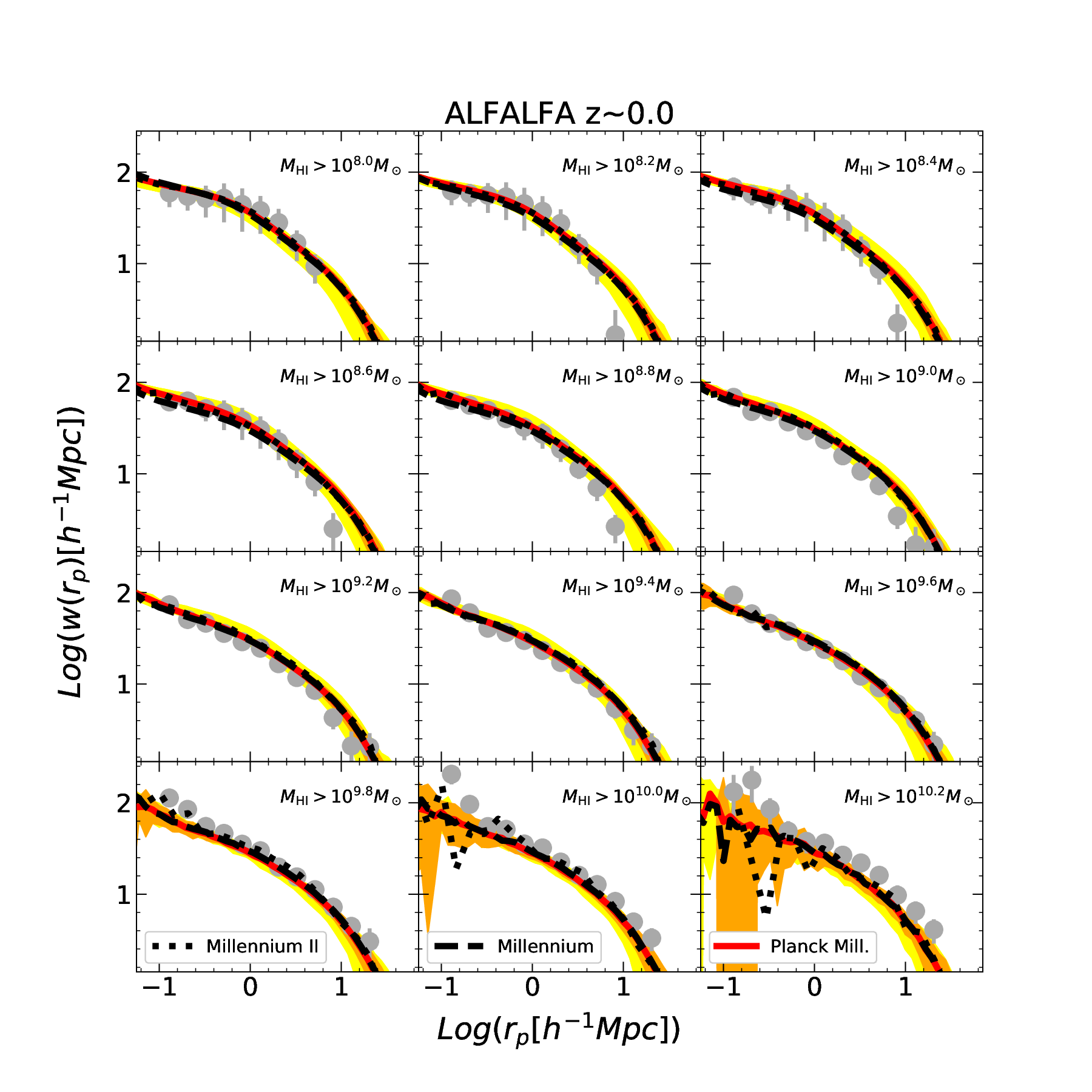}
  \caption{Projected 2-point correlation function for HI selected
    galaxies z$\lesssim$0.05; data from \citet{Guo17}. \gaea
    predictions refer to the realizations run on the MS (black dashed
    line) and on the \pms (red solid line) as in
    Fig.~\ref{fig:allconvA}.}\label{fig:wpgas}
\end{figure*}

The comparison presented in Fig.~\ref{fig:wpcol} indicates that \gaea
realizations reproduce qualitatively the relative clustering amplitude
between the red and blue populations. In order to better highlight
this we consider in Fig.~\ref{fig:wprat} the ratio between the
clustering amplitudes of the red and blue population at different
redshifts for some of the available mass bins. Fig.~\ref{fig:wprat}
shows that GAEA predictions reproduce qualitatively the observed
trends, with deviations due to the discrepancies discussed in relation
to Fig.~\ref{fig:wpcol}. The results presented in this section suggest
that the treatment for the physical mechanisms responsible for
quenching in \gaea not only correctly reproduces the overall assembly
and evolution of the quenched galaxy population (DL24,
\citealt{Xie24}), but it is also broadly consistent with their
expected distribution in the Large Scale Structure. Therefore, this
analysis represents a further positive test suggesting that \gaea
correctly captures the relative role of internal and environmental
processes in shaping the galaxy properties in different environments.

\subsection{2-point correlation function for HI selected galaxies}

In the previous sections, we study the 2pCF of galaxies selected on
the basis of their stellar content. A complementary analysis involves
the study of the cold gas content of galaxies, through the
distribution of HI selected galaxies. Gas distribution in galaxies at
low redshift can be mapped using the 21 cm HI hyper-fine emission
line. In particular, the Arecibo Fast Legacy ALFA Survey (ALFALFA;
\citealt{Giovanelli05}) samples more than 5000 galaxies out to
z$\sim$0.05, enabling the study of their clustering properties
\citet{Papastergis13}, showed that HI selected galaxies exhibit a
lower $w(r_p)$ amplitude with respect to stellar mass selected
samples. \citet{Guo17} used the 70\% complete ALFALFA catalogue and
showed that the 2pCF of of HI-selected galaxies depends strongly on
the HI mass, i.e. galaxies with higher HI reservoirs are more strongly
clustered on scales above a few Mpc than their counterparts with lower
HI content. We compare \citet{Guo17} estimates ($\pi_{\rm max}$=20
$\hmsun$) at different HI-mass (M$_{\rm HI}$) thresholds with \gaea
predictions in Fig.~\ref{fig:wpgas}; we find a general agreement in
the clustering amplitude and, most importantly, in its relative
evolution as a function of M$_{\rm HI}$. The larger discrepancies are
seen at the high-M$_{\rm HI}$ end, although data and predictions are
consistent within the statistical errorbars. This result is similar to
those based on M$_\star$ selected galaxies, although the reference
samples refer to different model galaxy populations, as most massive
galaxies at $z\sim0$ in \gaea have low gas fractions.

\section{Discussion and conclusions}
\label{sec:discconcl}

In this paper, we present predictions of the latest version of the
\gaea model, run on merger trees extracted from the \pms
simulation. We compare model results against a compilation of
observational measurements of galaxy clustering as a function of
stellar mass and star formation rate, covering a wide redshift range
from the local Universe up to the highest redshift when this
measurement has been possible (i.e. z$\sim$3.5). Our model shows a
satisfactory agreement with the observed projected 2pCF, of comparable
quality with respect to independent semi-analytic codes ({\sc
  L-galaxies} \citealt{Guo11, Henriques17}) and hydro-dynamical
simulations \citep{Springel18, Artale17}. At z$\sim$0, \gaea results
are close to those predicted from the {\sc TNG} hydro-simulation suite
\citep{Springel18}: a quite good agreement with data is found in the
10$^9 \msun$ < M$_\star$ < 10$^{11} \msun$ stellar mass range and at
the smallest scales, while the predicted 2pCT slightly under-predicts
observational measurements for the most massive galaxies. Similar
results have been also found for the {\sc EAGLE} simulation
\citep{Artale17}. Predictions from the {\sc L-galaxies} model are
closer to the observed $w(r_p)$ amplitude in the highest-mass bin, but
they either overpredict \citep{Guo11} or underpredict
\citep{Henriques17} the clustering of intermediate-to-low mass
galaxies (i.e.  10$^9 \msun \lesssim $ M$_\star \lesssim$ 10$^{10}
\msun$) At the lowest mass bin, our model predictions over-predict the
clustering signal, similarly to what found by \citet{Guo11}. Overall,
our model does not predict a significant evolution of the clustering
strength with redshift, in good agreement with independent theoretical
predictions \citep{Artale17} and observational measurements up to
z$\sim$1 \citep{Marulli13}. At higher redshift, data seem to suggest a
more significant evolution of the clustering amplitude than model
predictions, resulting in an overprediction with respect to the
observed normalization of the 2pCF. Considering the observational
uncertainties, model predictions are still consistent with
observational data, and more data at z$\gtrsim$2 are needed to clarify
if the highlighted discrepancy is significant.
\begin{figure}
  \includegraphics[width=9cm]{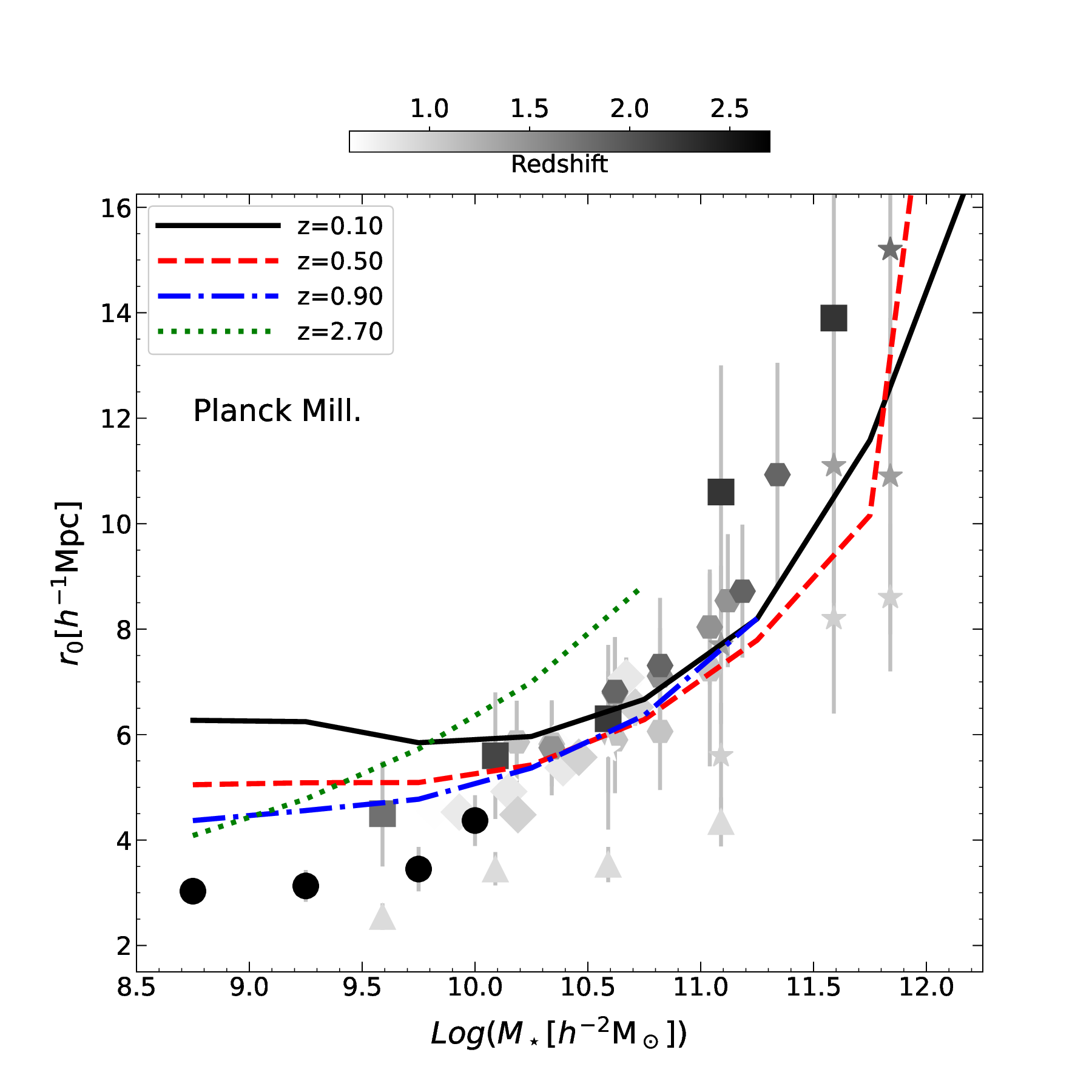}
  \caption{Clustering length $r_0$ as a function of stellar
    mass. \gaea predictions for the \pms at different redshifts are
    marked by different colours and linetypes as in the legend. The
    data compilation comes from \citet[triangles]{Meneux08},
    \citet[stars]{Foucaud10}, \citet[hexagons]{Wake11},
    \citet[squares]{Lin12}, \citet[diamonds]{Marulli13},
    \citet[circles]{Durkalec18}.}\label{fig:r0evo}
\end{figure}

Fig.~\ref{fig:r0evo} shows the dependence of the clustering lenght
$r_0$ on galaxy stellar mass and its redshift evolution as predicted
by our runs, and compares it to observational measurements. We compute
$r_0$ from the intrinsic three-dimensional 2pCF ($\xi$), directly
computed from our model outputs, defined as the scale corresponding to
$\xi(r_0)=1$. \gaea predicts a significant increase of $r_0$ as a
function of stellar mass.  For M$_\star>$10$^{10} \msun$ the predicted
$r_0$ values are in reasonable agreement with observational
constraints (see e.g. \citealt{Foucaud10, Wake11, Lin12, Marulli13}),
with a clear excess of the predicted value of r0 for galaxy stellar
masses below $\sim$10$^{10} \msun$. At fixed stellar mass, we find
only a weak evolution of $r_0$ at $z<2$, in agreement with the
numerical results from \citet{Artale17} and \citet{Springel18}. A
significant discrepancy between model predictions and data is found at
z$\gtrsim$2, with \gaea predicting a factor of $\sim$2 increase of
$r_0$ at fixed stellar mass, and data exhibiting a slower trend as a
function of galaxy stellar mass evolution.

If we split our model galaxies in star-forming and quiescent samples,
we find that \gaea correctly reproduces the increased clustering of
red quiescent galaxies with respect to blue star forming systems.
Overall, the red population shows a better agreement with data over
the entire galaxy mass range considered, while blue galaxies exhibit
both an excess of the clustering signal for low stellar masses
(M$_\star<$10$^{9}$ $h^{-1}$ M$_\odot$) and an underprediction of the
2pCF at large masses (M$_\star>$10$^{11}$ $h^{-1}$ M$_\odot$). Both
discrepancies are consistent with the tensions we find between model
prediction and data for the global population. In general, \gaea
results align well with similar comparisons shown in \citet{Guo11} and
\citet{Springel18} for an independent semi-analytic and
hydro-dynamical model of galaxy formation. These models tend to better
reproduce the properties of the star-forming population, while
overpredicting the clustering of the red population at intermediate
stellar mass (in particular, for M$_\star \sim$ 10$^{9}$-10$^{10}$
$h^{-1}$ M$_\odot$).

Finally, we also compare our model predictions with clustering
measurement for HI-selected galaxies in the ALFA survey and we find a
good agreement over the entire $M_{\rm HI}$ mass range considered.
Overall, the results discussed in this paper demonstrate that the
proposed modelling of physical mechanisms responsible for gas
consumption, recycling and quenching in \gaea (both internal and
environmental) are able to capture the main trends found for the
observed clustering strength and its variation with stellar mass, SF
activity and HI content. We stress that our model has not been
calibrated to reproduce galaxy clustering, that thus represents a
genuine model prediction: this is not a trivial success, and it shows
that our approach is not only able to recover the global properties of
galaxy populations in a statistical sense, but also their spatial
distribution in the LSS. The general good agreement between model
predictions and data for the 1-halo term (i.e. at scales $\lesssim$2-3
Mpc) for stellar masses down to M$_\star \sim$ 10$^{9}$ $h^{-1}$
M$_\odot$, suggests that the satellite distribution within halos is
correctly reproduced, at least down to this stellar mass limit. This
implies that our treatment for the evolution of satellite galaxies as
in \citet{Xie20} is generally effective in recovering the main
properties of this population, both in terms of stellar mass and star
formation activity. The excess of clustering signal found for lower
stellar masses can be understood as a combination of different
effects. On the one hand, satellite galaxies residing in low-mass
haloes are over-massive with respect to what is found in subhalo
abundance matching models constrained to reproduce both the 2pCF and
the GSMF \citep{Wang13b}. The clustering signal excess for the 2-halo
term ((i.e. at scales $\gtrsim$2-3 Mpc) suggests that \gaea is
predicting a larger than expected number of satellites in massive
haloes and this can be related to our treatment of the merging times
for orphan galaxies.

In summary, the latest \gaea version shows a remarkable level of
agreement for model predictions run on mergers trees extracted from
simulations of different resolution and (slightly) different
$\Lambda$CDM cosmological parameters. This result is driven by the
interplay between the new treatment for AGN (and its feedback on the
host galaxies) and the explicit partitioning of cold gas into its
atomic and molecular components. Both physical mechanisms act as
regulators of the cold gas in model galaxies. \gaea reasonably
reproduces the observed dependence of galaxy clustering on stellar
mass and star formation activity. Moreover, model predictions are also
broadly consistent with the clustering measurements corresponding to
HI-selected galaxies. This is an important confirmation that our
treatment of the mechanisms acting on the baryonic component of
galaxies, and satellites in particular, is able to capture the
relevant physical dependencies. \gaea predicts a small redshift
evolution of the amplitude of the 2pCF: this is consistent with
available data up to z$\sim$3.5, which we interpret as another
indication that our model is able to correctly connect galaxies with
their environment. At higher redshifts, our model tends to
under-predict the observational measurements. However, larger
observational samples are needed to draw firmer conclusions in this
redshift regime.

\begin{acknowledgements}
An introduction to {\gaea}, a list of our recent work, as well as
datafile containing published model predictions, can be found at
\url{https://sites.google.com/inaf.it/gaea/home}.
  
We acknowledge the use of INAF-OATs computational resources within the
framework of the CHIPP project \citep{Taffoni20} and the INAF PLEIADI
program (\url{http://www.pleiadi.inaf.it}). This work used the
DiRAC@Durham facility managed by the Institute for Computational
Cosmology on behalf of the STFC DiRAC HPC Facility
(\url{www.dirac.ac.uk}). The equipment was funded by BEIS capital funding
via STFC capital grants ST/P002293/1, ST/R002371/1 and ST/S002502/1,
Durham University and STFC operations grant ST/R000832/1. DiRAC is
part of the National e-Infrastructure.

We thank the anonymous referee for a careful reading of the manuscript
that helped us improve the quality of the presentation of our
results. We thank Chen Li for sharing his observational estimates for
the clustering signal in the SDSS and DEEP2 surveys. We acknowledge
stimulating discussions with Andrea Biviano, Alessandra Fumagalli,
Federico Marulli, Pierluigi Monaco and Emiliano Sefusatti. MH
acknowledges funding from the Swiss National Science Foundation (SNSF)
via a PRIMA grant PR00P2-193577 ‘From cosmic dawn to high noon: the
role of BHs for young galaxies’.

\end{acknowledgements}

\bibliographystyle{aa} 
\bibliography{fontanot} 
\begin{appendix} 
\label{sec:app}
\section{Additional model predictions}
\begin{figure}
    \includegraphics[width=9cm]{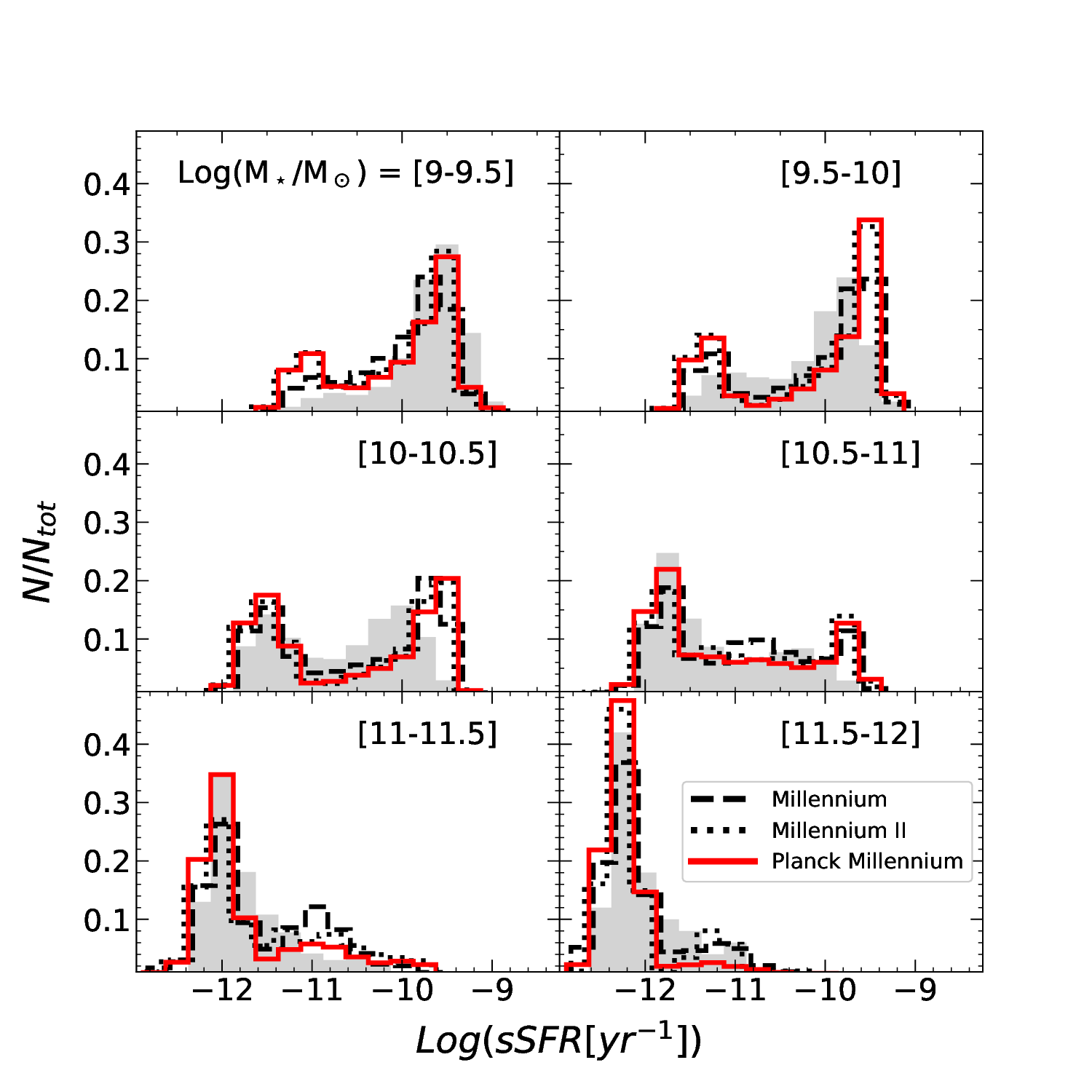}
  \caption{Specific star formation rate distribution in stellar mass
    bins, compared to SDSS DR8 observational estimates (grey shaded
    histograms). MS and MSII histograms have been slightly shifted (by
    0.05 dex) for clarity. Black solid, black dashed and red solid
    lines refer the \gaea model version presented in DL24 and run on
    the MS, MSII and \pms, respectively.}\label{fig:allconvD}
\end{figure}
\begin{figure}
    \includegraphics[width=9cm]{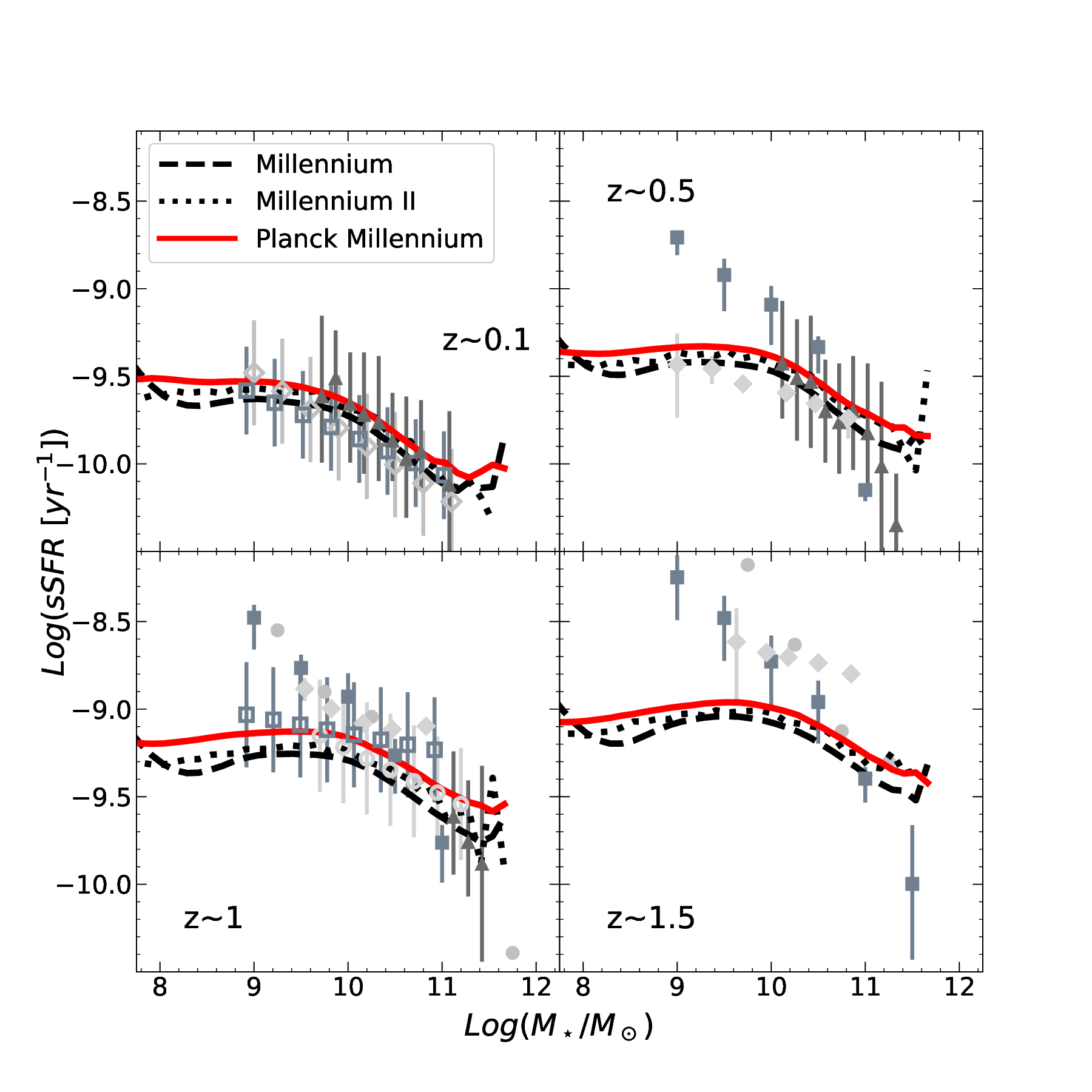}
  \caption{Redshift evolution of the main sequence of star-forming
    galaxies. Line-types and colours as in Fig.~\ref{fig:allconvD},
    datapoints correspond to the compilations used in \citet[][filled
      symbols, see complete reference list for their
      Fig.~5]{Fontanot09b} and in \citet[][empty symbols, see complete
      reference list for their Fig.~11]{Xie17}.}\label{fig:allconvF}
\end{figure}
\begin{figure}
    \includegraphics[width=9cm]{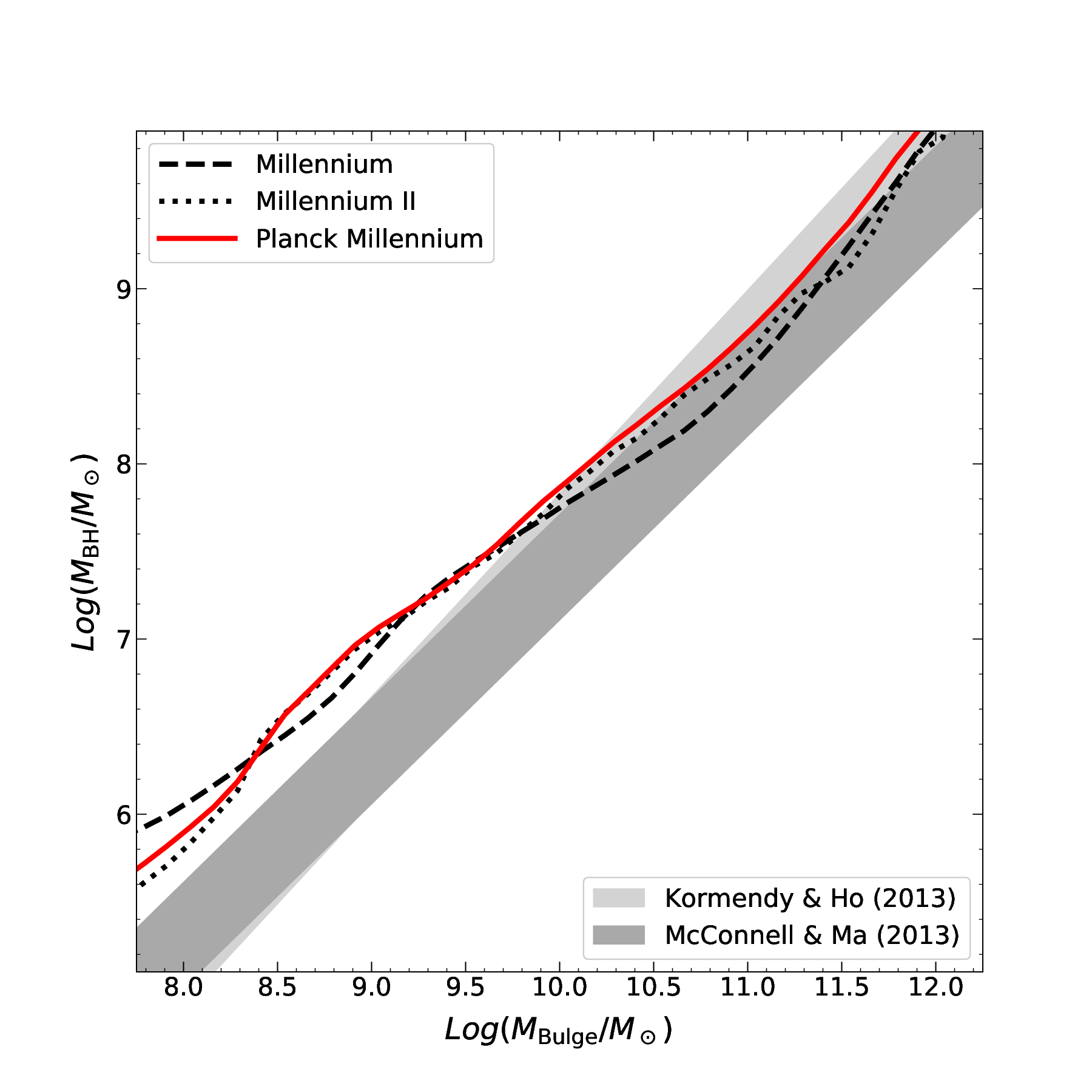}
  \caption{M$_{\rm BH}$-M$_{\rm Bulge}$ scaling relation at
    z$\sim$0. Shaded areas correspond to the best fit relations
    of~\citet{KormendyHo13} and~\citet{McConnellMa13}, for both
    relations we assume a scatter of 0.3 dex. Line-types and colours
    as in Fig.~\ref{fig:allconvD}. }\label{fig:allconvC}
\end{figure}
\begin{figure*}
  \centerline{
    \includegraphics[width=9cm]{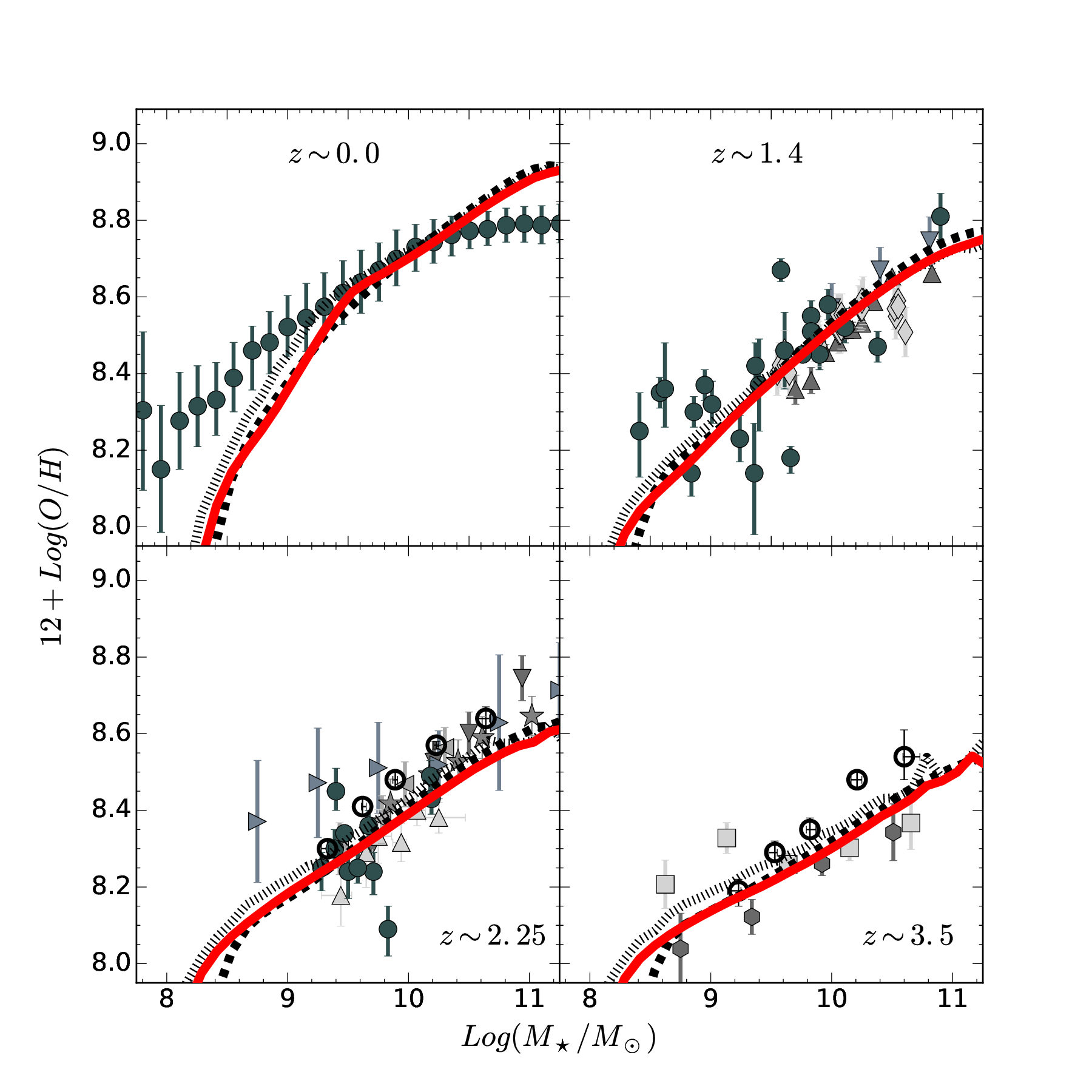}
    \includegraphics[width=9cm]{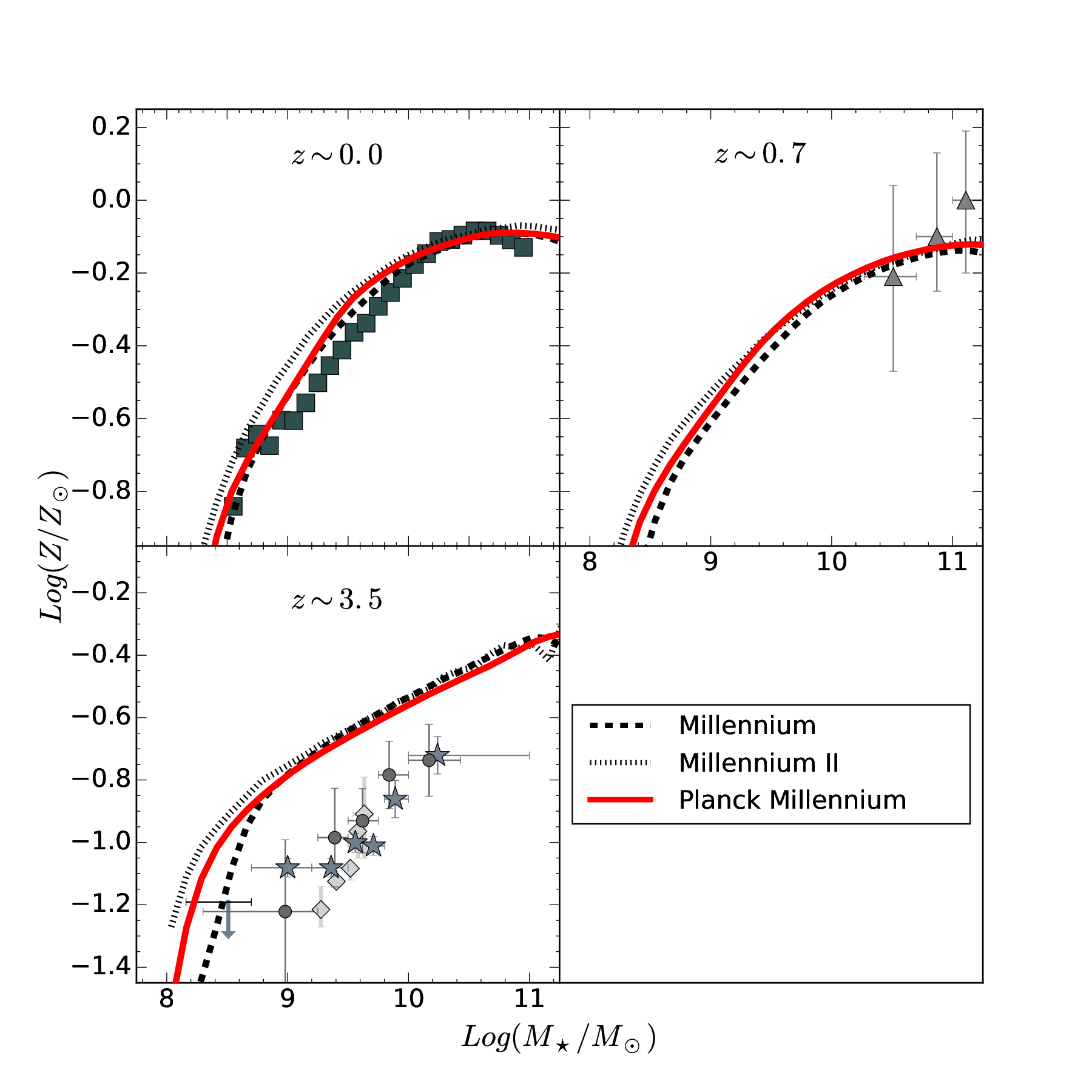} }
  \caption{{\it Left panel:} Redshift evolution of the cold gas
    metallicity as a function of stellar mass (following
    \citealt{Fontanot21} we apply a downwards 0.1 dex shift to the
    intrinsic predictions). {\it Right panel:} stellar
    mass-metallicity relation at different redshifts. Line-types and
    colours as in Fig.~\ref{fig:allconvD}; in both panels datapoints
    correspond to the compilation used in \citet[][see complete
      reference list for their
      Fig.~1]{Fontanot21}.}\label{fig:allconvE}
\end{figure*}
\begin{figure*}
  \centerline{ \includegraphics[width=18cm]{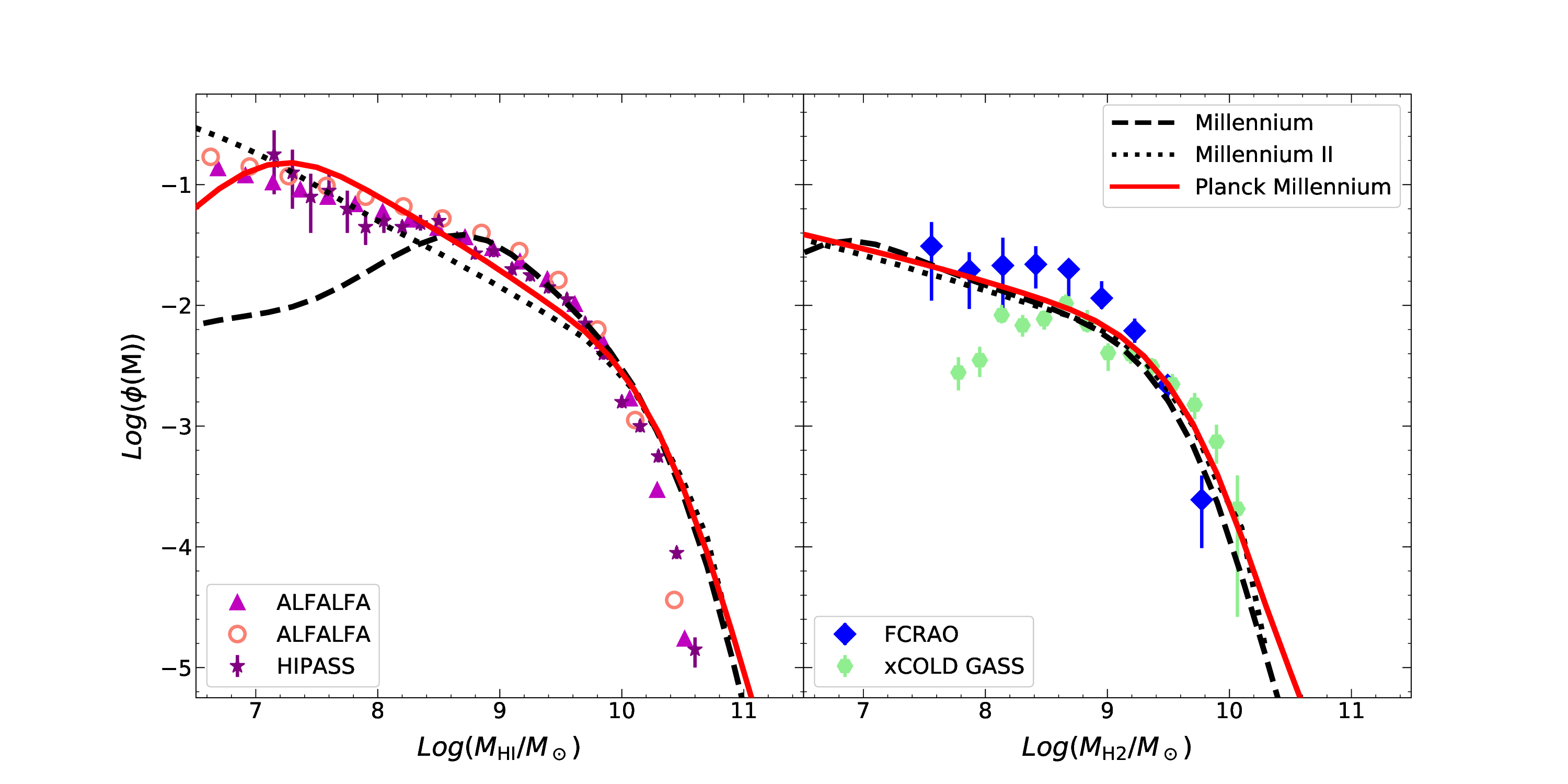} }
  \caption{HI and H2 mass functions at z$\sim$0 are shown in the left
    and right panel respectively. Observational data are from
    \citet[][blue empty diamonds]{Keres03}, \citet[][purple
      stars]{Zwaan05}, \citet[][magenta triangles]{Haynes11},
    \citet[][pink empty circles]{Jones18} and \citet[][light green
      hexagons]{Fletcher21}. Line-types and colours are as in
    Fig.~\ref{fig:allconvD}. Model predictions have been convolved
    with an estimated 0.25 dex error on the gas mass
    determination.}\label{fig:allconvB}
\end{figure*}

In this Appendix we show additional observational constraints to
further explore the convergence of the latest \gaea model, and to
demonstrate that our previously published results hold in the latest
rendition of our model. We consider in Fig.~\ref{fig:allconvD} the
z$\sim$0 distribution of sSFR in stellar mass bins and in
Fig.~\ref{fig:allconvF} the evolution of the main sequence of star
forming galaxies. In the latter figure, model star-forming galaxies
have been selected as those with sSFR larger than 0.3/t$_H$, where
t$_H$ is the Hubble time at the redshift under
consideration. Fig.~\ref{fig:allconvC} shows the z$\sim$0 scaling
relation between the mass of the central SMBH (M$_{\rm BH}$) and the
mass of the spheroidal component (M$_{\rm Bulge}$), while
Fig.~\ref{fig:allconvE} shows the gas-phase and stellar
mass-metallicity relations. In Fig.~\ref{fig:allconvB}, we report the
mass functions at $z=0$ for the neutral and molecular gas
components. These distributions have been used for the calibration of
the model on MS merger trees \citep{Xie17,
  DeLucia24}. Fig.~\ref{fig:allconvB} shows that for the HI and H2
mass function the consistency between the model realizations run on
different simulations is higher (i.e. it extends to smaller gas
masses) than previous versions of the model (see e.g. Fig.~7 in
\citealt{Xie17}).

\end{appendix}

\end{document}